\documentclass[aps,reprint]{revtex4-1}
\usepackage{amsfonts}
\usepackage{amsmath}
\usepackage{graphicx}
\usepackage[colorlinks]{hyperref}
\usepackage[utf8]{inputenc}
\usepackage{tikz}
\usetikzlibrary{positioning}
\begin{document}
\hypersetup{%
  citecolor=[rgb]{0.1804,0.1882,0.5725},%
  urlcolor=[rgb]{0.1804,0.1882,0.5725},%
  linkcolor=[rgb]{0.1804,0.1882,0.5725}%
}%

\title{Automated Discovery of Jet Substructure Analyses}

\author{Yue Shi Lai}

\affiliation{Lawrence Berkeley National Laboratory, Berkeley, CA
  94720, USA}

\begin{abstract}
The study of the substructure of collimated particles from quarks and
gluons, or jets, has the promise to reveal the details how color
charges interact with the QCD plasma medium created in colliders such
as RHIC and the LHC. Traditional jet substructure observables have
been constructed using expert knowledge, and are largely transplanted,
unmodified, from the high-energy physics, where the goal is primarily
the study of boosted hadronic decays. A novel neural network
architecture is described that is capable of examining theoretical
models, and constructs, on its own, an analysis procedure that is
sensitive to the internal model features. This architecture, in
combination with symbolic regression, further allows the extraction of
closed-form algebraic expressions from the learned result -- enabling
the automatically constructed jet substructure analysis to be
subsequently understood and reproduced by humans. This system is then
tasked to construct an analysis that infers the plasma temperature
from observing jets, which is demonstrated using both JEWEL and the
Linearized Boltzmann Transport model, and at the presence of a
realistic remnant of the plasma, or underlying event, that the
measurement has to overcome. In a demonstration how algorithms can
produce original research in direct competition to human experts, the
resulting jet substructure variables and analyses are capable of
determining the initial temperature of the plasma medium from
analyzing 1200--2500 jets, a performance not seen in existing,
manually designed analyses. Comparison of an incidentally discovered
observable with the existing literature further indicates that the
system described is capable of examining the model phase spaces to a
detail at least comparable to the current field of human experts.
\end{abstract}

\maketitle

Colliders, such as the Large Hadron Collider (LHC) at CERN, can be
used to heat nuclei to very high temperature and compress them to
densities many times that of normal nuclei. It has been shown that
heavy ion collisions at both RHIC and the LHC undergo a phase
transition from normal, bound hadronic matter to a plasma of quarks
and gluons. This quark--gluon plasma has surprising properties: it
flows as a nearly frictionless fluid, and exhibits a large opacity to
transiting quarks and
gluons~\cite{doi:10.1146/annurev-nucl-102212-170540,*ADCOX2005184,*%
ADAMS2005102}.

Analogous to the Bethe formula known for the electromagnetic charge
and plasma, a key question is the magnitude and mechanism of energy
loss by quarks and gluons (partons) passing through quark--gluon
plasma, and how the plasma transports the deposited energy. Addressing
this experimentally requires observables sensitive to the interaction
between partons and the plasma. A novel approach uses jet substructure
observables, built from the angular correlation of energies inside the
collimated spray of hadrons (known as a jet) that a parton becomes
before reaching the detector.

In heavy ion publications utilizing jet substructure
(e.g.\ \cite{Mehtar-Tani2017,*PhysRevLett.119.112301,*%
  KunnawalkamElayavalli2017,*PhysRevLett.120.142302}), the
substructure variables re-use those developed to tag boosted objects
in high energy physics. So far, most known substructure analyses are
moderately sensitive to the presence of a heavy-ion collision vs. the
proton--proton baseline, but do not demonstrate a sensitivity to
specific heavy ion model features. Consequently, we do not know
whether jet substructure can provde as much information about quark
gluon plasma properties as existing measurements of the soft, bulk
emission~\cite{PhysRevLett.106.192301,%
  *PhysRevC.83.044911,*PhysRevC.83.054912,*PhysRevLett.110.012302}. In
this article, examples of novel jet substructure variables are given,
together with a neural network (NN) based method that led to their
discovery. The demonstration of such an automatically produced,
previously unknown result is also a demonstration how algorithms can
produce original research in direct competition to human experts.

Possible analyses that can be applied to Quantum Chromodynamics (QCD)
in hadron or heavy ion collider experiments can be expressed as a
combination of two functions, the per-event observable extraction and
a subsequent statistical analysis. A human expert would construct
analyses iteratively via generating hypothesis from his or her
knowledge or intuition, and testing it against models. However, when
the function space of possible analyses is very large, a competing
method would be automated search for an analysis of the desired
property, using numerical optimization. Neural Networks are known as
universal function approximators~\cite{HORNIK1989359,*LESHNO1993861},
and deep layered NN have been demonstrated to be more efficient than
traditional ``shallow'' function approximation
techniques~\cite{MhaskarP16,*LiangS16,*YAROTSKY2017103}.

First, the general formulation of analysis functions using the
structure of the NN is described. Then, taking advantage of the
efficient optimization that can be applied to NN, analyses are
constructed by optimizing performance extracting physics parameters
from Monte Carlo models, without involving human physics knowledge in
designing the analysis. This is in departure from previous instances
of automated generation of scientific hypotheses for research, where
(non-mathematical) domain-specific, human knowledge are used as
input~(e.g.\ \cite{Zytkow:1990:ADC:1865609.1865633,*LINDSAY1993209,*%
  King2004,*VOYTEK201292,*72d570723c644a8cbb07011a2d39d526,*%
  Spangler:2014:AHG:2623330.2623667,*%
  doi:10.1021/acs.molpharmaceut.7b00346,*PhysRevLett.116.090405}).

Events for the NN training are generated for lead--lead (Pb-Pb)
collision at a center-of-mass energy of $\sqrt{s_{NN}} =
5.02\:\mathrm{TeV}$, corresponding to the Large Hadron Collider (LHC)
Run-2 data. The impact parameter range sampled corresponds to the
$0$--$10\%$ most central collision geometries among the total
inelastic Pb-Pb cross section from the Glauber Monte Carlo in
\cite{PhysRevC.88.044909}. Jets that interact with the plasma medium
are generated using \textsc{Jewel} 2.2.0~\cite{Zapp2014} and the
Linearized Boltzmann Transport (LBT) model~\cite{PhysRevC.91.054908}.
Events are weighted $\propto \hat{p}_\perp^{5.7}$, where
$\hat{p}_\perp$ is the center-of-mass transverse momentum transfer in
a single parton-parton scattering. This approximately compensates for
the power law decrease of the jet spectrum with jet energy.

\textsc{Jewel} events are generated with its default plasma model,
where the initial time $\tau_\mathrm{i}$ is varied between $0.2$ and
$0.8\:\mathrm{fm}/c$, and the mean initial temperature $T_\mathrm{i}$
between $0.16$ and $0.76\:\mathrm{GeV}/k$. LBT events are generated
using parton level hard scattering from \textsc{pythia}
8.235~\cite{SJOSTRAND2015159} tune CUETP8M1~\cite{Khachatryan2016}.
Recoiling scattering centers are subtracted according to the procedure
the \textsc{Jewel} author has referred to as ``4MomSub'', which
clusters those medium partons into the jet using near-zero momentum
``ghost particles'' that are place holders and then subtract their
original four-momenta from the jet substructure.

For LBT, in order to study the sensitivity to initial parameters
without costly rerunning of numerical hydrodynamics, the temperature
and velocity profile is sampled from the viscous Gubser
flow~\cite{PhysRevD.78.066014,*PhysRevD.82.085027} as its plasma
dynamics model. The integration constant $\hat{T}_0$ is varied between
$0.383$ and $0.583\:\mathrm{GeV}/k$, such that the resulting medium
temperature matches with $T(\tau_\mathrm{i}) = T_\mathrm{i}$ the same
range as \textsc{Jewel}. As jets in LBT interact with the medium
exchange color charges with the recoiling medium partons, the hard
scattering event contains color connection with medium partons that
are not actually present in the final state. In order to hadronize the
event in \textsc{pythia} 8, those medium partons are retained, and are
added back to the hard scattering event with zero momenta. The same
procedure as ``4MomSub'' for \textsc{Jewel}, as described previously,
is also applied here.

\begin{figure}
  \begin{tikzpicture}[scale=0.65,every node/.style={minimum size=0.03125in},on grid]
  \begin{scope}[
      xshift=300,every node/.append style={
        xscale=0.5,yslant=-0.866},xscale=0.5,yslant=-0.866
    ]
    \draw (1.5,2.75) coordinate (a6);
    \draw (1.5,2.25) coordinate (b6);
    \draw (2,2.25) coordinate (c6);
    \draw (2,2.75) coordinate (d6);
    \fill[white,fill opacity=0.6] (1.5,2.25) rectangle (2.5,2.75);
    \draw[step=0.5,black,shift={(1.5,2.25)}] (0,0) grid (1,0.5);
    \draw[black,very thick,line join=round] (1.5,2.25) rectangle (2.5,2.75);
    \fill[gray] (1.6,2.35) rectangle (1.9,2.65);
  \end{scope}
  \begin{scope}[
      xshift=245,every node/.append style={
        xscale=0.5,yslant=-0.866},xscale=0.5,yslant=-0.866
    ]
    \draw (0.5,4) coordinate (a5);
    \draw (0.5,1) coordinate (b5);
    \draw (3.5,1) coordinate (c5);
    \draw (3.5,4) coordinate (d5);
    \draw[black] (a5) -- (b5) -- (b6) -- (a6) -- cycle;
    \draw[black] (b5) -- (c5) -- (c6) -- (b6) -- cycle;
    \fill[white,fill opacity=0.6] (a5) -- (d5) -- (d6) -- (a6) -- cycle;
    \fill[white,fill opacity=0.6] (d5) -- (c5) -- (c6) -- (d6) -- cycle;
    \draw[black] (a5) -- (d5) -- (d6) -- (a6) -- cycle;
    \draw[black] (d5) -- (c5) -- (c6) -- (d6) -- cycle;
    \draw (0.5,4) coordinate (a5);
    \draw (0.5,3.5) coordinate (b5);
    \draw (1,3.5) coordinate (c5);
    \draw (1,4) coordinate (d5);
    \fill[white,fill opacity=0.6] (0.5,1) rectangle (3.5,4);
    \draw[step=0.5, black] (0.5,1) grid (3.5,4);
    \draw[black,very thick,line join=round] (0.5,1) rectangle (3.5,4);
    \foreach \i in {0,...,5} {
      \foreach \j in {0,...,5} {
        \fill[gray] (0.5 * \i + 0.6, 0.5 * \j + 1.1)
        rectangle (0.5 * \i + 0.9, 0.5 * \j + 1.4);
      }
    }
  \end{scope}
  \begin{scope}[
      xshift=197.5,every node/.append style={
        xscale=0.5,yslant=-0.866},xscale=0.5,yslant=-0.866
    ]
    \draw (0.5,4) coordinate (a4);
    \draw (0.5,1) coordinate (b4);
    \draw (3.5,1) coordinate (c4);
    \draw (3.5,4) coordinate (d4);
    \draw[black] (a4) -- (b4) -- (b5) -- (a5) -- cycle;
    \draw[black] (b4) -- (c4) -- (c5) -- (b5) -- cycle;
    \fill[white,fill opacity=0.6] (a4) -- (d4) -- (d5) -- (a5) -- cycle;
    \fill[white,fill opacity=0.6] (d4) -- (c4) -- (c5) -- (d5) -- cycle;
    \draw[black] (a4) -- (d4) -- (d5) -- (a5) -- cycle;
    \draw[black] (d4) -- (c4) -- (c5) -- (d5) -- cycle;
    \draw (0.5,4) coordinate (a4);
    \draw (0.5,3.5) coordinate (b4);
    \draw (1,3.5) coordinate (c4);
    \draw (1,4) coordinate (d4);
    \fill[white,fill opacity=0.6] (0.5,1) rectangle (3.5,4);
    \draw[step=0.5, black] (0.5,1) grid (3.5,4);
    \draw[black,very thick,line join=round] (0.5,1) rectangle (3.5,4);
    \foreach \i in {0,...,5} {
      \foreach \j in {0,...,5} {
        \fill[gray] (0.5 * \i + 0.6, 0.5 * \j + 1.1)
        rectangle (0.5 * \i + 0.9, 0.5 * \j + 1.4);
      }
    }
  \end{scope}
  \begin{scope}[
      xshift=150,every node/.append style={
        xscale=0.5,yslant=-0.866},xscale=0.5,yslant=-0.866
    ]
    \draw (0.5,4) coordinate (a3);
    \draw (0.5,1) coordinate (b3);
    \draw (3.5,1) coordinate (c3);
    \draw (3.5,4) coordinate (d3);
    \draw[black] (a3) -- (b3) -- (b4) -- (a4) -- cycle;
    \draw[black] (b3) -- (c3) -- (c4) -- (b4) -- cycle;
    \fill[white,fill opacity=0.6] (a3) -- (d3) -- (d4) -- (a4) -- cycle;
    \fill[white,fill opacity=0.6] (d3) -- (c3) -- (c4) -- (d4) -- cycle;
    \draw[black] (a3) -- (d3) -- (d4) -- (a4) -- cycle;
    \draw[black] (d3) -- (c3) -- (c4) -- (d4) -- cycle;
    \draw (3,3.5) coordinate (a3);
    \draw (3,3) coordinate (b3);
    \draw (3.5,3) coordinate (c3);
    \draw (3.5,3.5) coordinate (d3);
    \fill[white,fill opacity=0.6] (0.5,1) rectangle (3.5,4);
    \draw[step=0.5, black] (0.5,1) grid (3.5,4);
    \draw[black,very thick,line join=round] (0.5,1) rectangle (3.5,4);
    \foreach \i in {0,...,5} {
      \foreach \j in {0,...,5} {
        \fill[gray] (0.5 * \i + 0.6, 0.5 * \j + 1.1)
        rectangle (0.5 * \i + 0.9, 0.5 * \j + 1.4);
      }
    }
  \end{scope}
  \begin{scope}[
      xshift=102.5,every node/.append style={
        xscale=0.5,yslant=-0.866},xscale=0.5,yslant=-0.866
    ]
    \draw (2.5,5) coordinate (a2);
    \draw (2.5,0) coordinate (b2);
    \draw (3.5,0) coordinate (c2);
    \draw (3.5,5) coordinate (d2);
    \draw[black] (a2) -- (b2) -- (b3) -- (a3) -- cycle;
    \draw[black] (b2) -- (c2) -- (c3) -- (b3) -- cycle;
    \fill[white,fill opacity=0.6] (a2) -- (d2) -- (d3) -- (a3) -- cycle;
    \fill[white,fill opacity=0.6] (d2) -- (c2) -- (c3) -- (d3) -- cycle;
    \draw[black] (a2) -- (d2) -- (d3) -- (a3) -- cycle;
    \draw[black] (d2) -- (c2) -- (c3) -- (d3) -- cycle;
    \draw (0.5,5) coordinate (a2);
    \draw (0.5,4.5) coordinate (b2);
    \draw (1,4.5) coordinate (c2);
    \draw (1,5) coordinate (d2);
    \fill[white,fill opacity=0.6] (0.5,0) rectangle (3.5,5);
    \draw[step=0.5, black] (0.5,0) grid (3.5,5);
    \draw[black,very thick,line join=round] (0.5,0) rectangle (3.5,5);
    \fill[gray] (0.6, 4.6) rectangle (0.9, 4.9);
    \foreach \i in {0,...,1} {
      \foreach \j in {0,...,9} {
        \fill[gray] (0.5 * \i + 2.6, 0.5 * \j + 0.1)
        rectangle (0.5 * \i + 2.9, 0.5 * \j + 0.4);
      }
    }
  \end{scope}
  \begin{scope}[
      xshift=55,every node/.append style={
        xscale=0.5,yslant=-0.866},xscale=0.5,yslant=-0.866
    ]
    \draw (0.5,5) coordinate (a1);
    \draw (0.5,4.5) coordinate (b1);
    \draw (3.5,4.5) coordinate (c1);
    \draw (3.5,5) coordinate (d1);
    \draw[black] (a1) -- (b1) -- (b2) -- (a2) -- cycle;
    \draw[black] (b1) -- (c1) -- (c2) -- (b2) -- cycle;
    \fill[white,fill opacity=0.6] (a1) -- (d1) -- (d2) -- (a2) -- cycle;
    \fill[white,fill opacity=0.6] (d1) -- (c1) -- (c2) -- (d2) -- cycle;
    \draw[black] (a1) -- (d1) -- (d2) -- (a2) -- cycle;
    \draw[black] (d1) -- (c1) -- (c2) -- (d2) -- cycle;
    \draw (0.5,5) coordinate (a1);
    \draw (0.5,4.5) coordinate (b1);
    \draw (1,4.5) coordinate (c1);
    \draw (1,5) coordinate (d1);
    \fill[white,fill opacity=0.6] (0.5,0) rectangle (3.5,5);
    \draw[step=0.5, black] (0.5,0) grid (3.5,5);
    \draw[black,very thick,line join=round] (0.5,0) rectangle (3.5,5);
    \foreach \i in {0,...,5} {
      \fill[gray] (0.5 * \i + 0.6, 4.6) rectangle (0.5 * \i + 0.9, 4.9);
    }
  \end{scope}
  \begin{scope}[
      xshift=0,every node/.append style={
        xscale=0.5,yslant=-0.866},xscale=0.5,yslant=-0.866
    ]
    \draw (0,5) coordinate (a0);
    \draw (0,4.5) coordinate (b0);
    \draw (4,4.5) coordinate (c0);
    \draw (4,5) coordinate (d0);
    \draw[black] (a0) -- (b0) -- (b1) -- (a1) -- cycle;
    \draw[black] (b0) -- (c0) -- (c1) -- (b1) -- cycle;
    \fill[white,fill opacity=0.6] (a0) -- (d0) -- (d1) -- (a1) -- cycle;
    \fill[white,fill opacity=0.6] (d0) -- (c0) -- (c1) -- (d1) -- cycle;
    \draw[black] (a0) -- (d0) -- (d1) -- (a1) -- cycle;
    \draw[black] (d0) -- (c0) -- (c1) -- (d1) -- cycle;
    \fill[white,fill opacity=0.6] (0,0) rectangle (4,5);
    \draw[step=0.5,black] (0,0) grid (4,5);
    \draw[black,very thick,line join=round] (0,0) rectangle (4,5);
    \foreach \i in {0,...,7} {
      \fill[gray] (0.5 * \i + 0.1, 4.6) rectangle (0.5 * \i + 0.4, 4.9);
    }
    \draw[thick,->] (-0.5,0) -- (-0.5,5) node[midway,above,rotate=90] {Jet--events};
    \draw[thick,->] (0,-0.5) -- (4,-0.5) node[midway,below] {Substructure feature};
  \end{scope}
  \draw (a0) -- (a1) node[midway,above] { conv1d };
  \draw (a1) -- (a2) node[midway,above] { conv1d };
  \draw (6.2,0.6) node { sym- };
  \draw (6.2,0.3) node { metriza- };
  \draw (6.2,0) node { tion };
  \draw (6.7,2.8) node { dense };
  \draw (a4) -- (a5) node[midway,above] { dense };
  \draw (a5) -- (a6) node[midway,above] { softmax };
  \draw (1.9,-4.5) node[right] { $\underbrace{\hskip 0.845in\relax}_{c(\cdot)}\underbrace{\hskip 0.4225in\relax}_{s(\cdot)}\underbrace{\hskip 1.17in\relax}_{d(\cdot)}$ };
\end{tikzpicture}
  \caption{The layout of the statistical analysis learning neural
    network, where the functions $c$, $s$, and $d$ correspond to the
    convolutional feature extraction, permutation symmetrization, and
    dense layers.}
  \label{fig:nn_layout}
\end{figure}
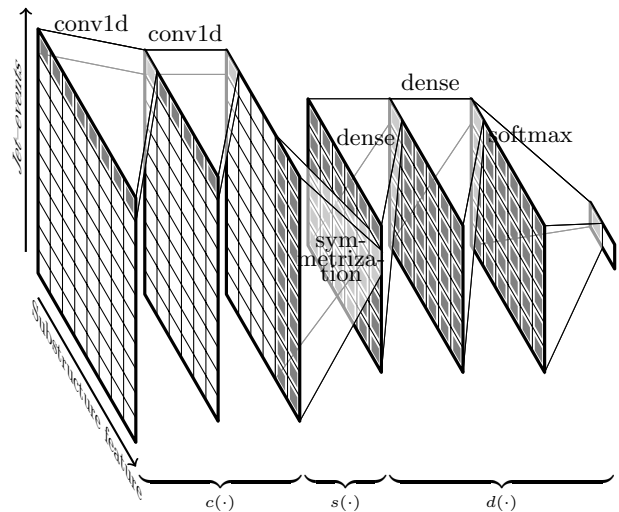

Neither \textsc{Jewel} nor LBT produces the particles that is the
remnant of the plasma medium, or the underlying event (UE). The UE,
for both \textsc{Jewel} and LBT, is generated using \textsc{hydjet}
1.9~\cite{Lokhtin2006}. Maintaining its default tune, the charged
particle multiplicity density $\langle dN_\mathrm{ch}/d\eta\rangle$
with the track pseudorapidity $\lvert\eta\rvert < 0.5$, at
$\sqrt{s_{NN}} = 5.02\:\mathrm{TeV}$ Pb-Pb, is observed to be $\approx
18\%$ higher ($2407\pm 5$ for $0$--$2.5\%$ centrality and $1787\pm 4$
for $7.5$--$10\%$ centrality) than experimentally
measured~\cite{PhysRevLett.116.222302}. Since the constructed analyses
must be robust in a more adverse UE environment than experimentally
encountered, the unmodified tune is sufficient for this study. Event
centralities are sampled independently in \textsc{Jewel} and
\textsc{hydjet}, in order to prevent the machine learned analysis to
be based on trivial multiplicity effects.

\begin{figure}
  \includegraphics[width=3.375in]{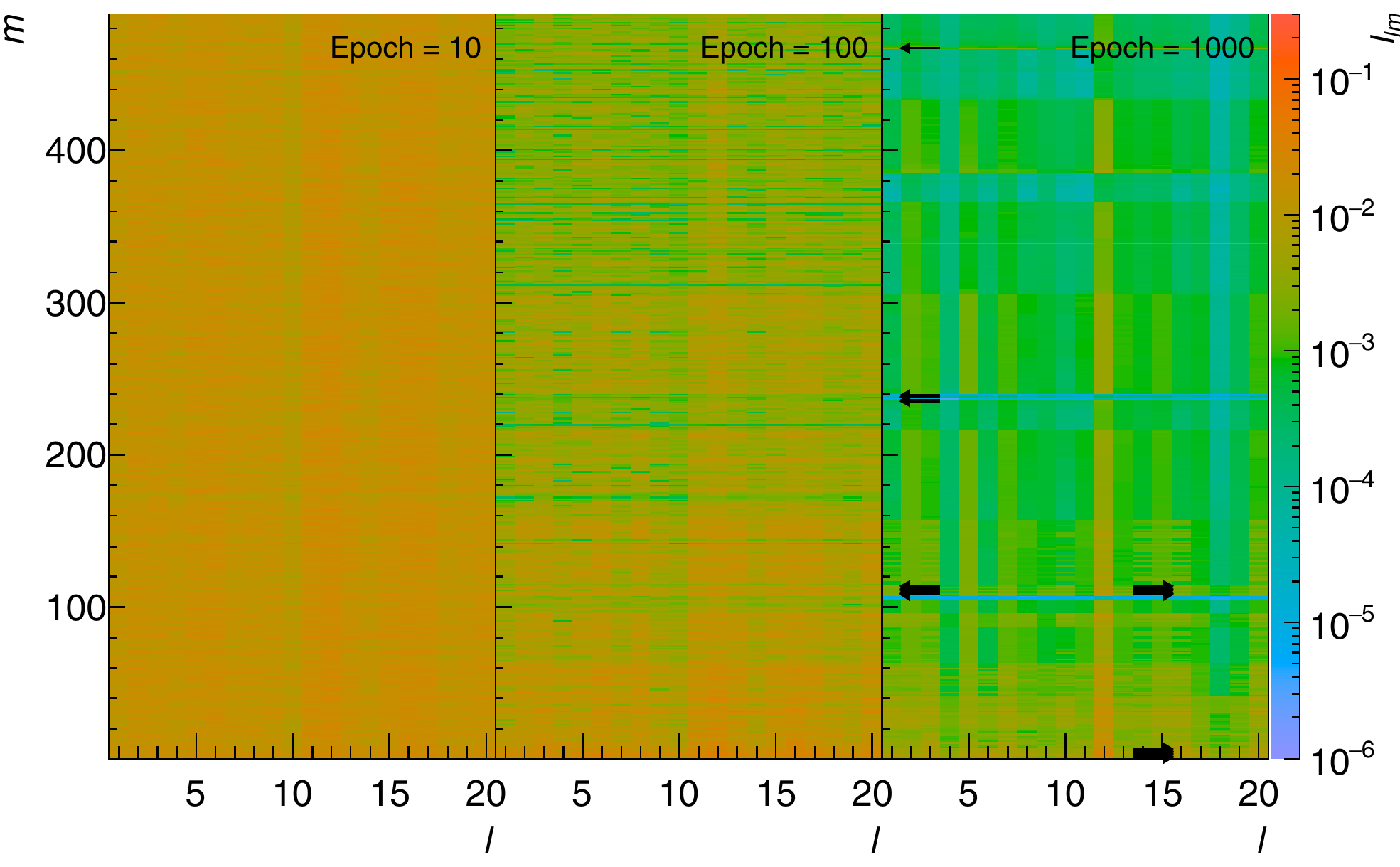}
  \caption{The upper bound on partial derivative $I_{lm}$, used as
    regularization, for the per jet--event analysis $c$ that
    discriminates between \textsc{Jewel} $T_\mathrm{i}$. The index $m
    \in \{1, 2, \ldots, 489\}$ counts the energy flow polynomial with
    $1\le\text{degree}\le 7$ (with increasing order), and $l$ is the
    index of the intermediate analysis. The progression is plotted for
    three different epochs, being full passes through the training
    data. Arrows in the last training epoch indicates the variable
    subsequently found in the approximation term found by symbolic
    regression (with $l = 1$ and $16$).}
  \label{fig:jewel_regularization}
\end{figure}

Jets are reconstructed using the anti-$k_T$
algorithm~\cite{1126-6708-2008-04-063} with the distance parameter $D
= 0.4$. Jet reconstruction is applied to the hard and UE event final
state particles superimposed, thus capturing the effect of imperfect
reconstruction due to the presence of the UE. A collision centrality
dependent mean UE particle contribution to the jet transverse momentum
$\langle p_{T,\mathrm{UE}}\rangle$ is determined. This is subtracted
to obtain the corrected $p_{T,J} = p_{T,\mathrm{tot}} - \langle
p_{T,\mathrm{UE}}\rangle$ at the hard collision scale. Jets with $100
< p_{T,J} < 500\:\mathrm{GeV}/c$ and $\lvert\eta_J\rvert < 1.4$ are
considered for the substructure analysis, ranges that are well-covered
by barrel tracking and calorimetry at the LHC. At these energies, jets
from hard scattering are reliably distinguished from the combinatorial
overlap of bulk-produced particles. The jet spectrum $dN_J/dp_{T,J}$
for different \textsc{Jewel} and LBT medium scenarios are forced to be
identical, by randomly discarding jets from the scenario with the
higher yield. This prevents the NN to produce a non-substructure
analysis that measures the jet spectrum. Also, the jets are analyzed
as if they are from independent events, since the aim is to observe
the effect of the medium on the substructure and not e.g.\ the
momentum balance.

\begin{figure*}
  \centerline{%
    \includegraphics[height=3.25in]{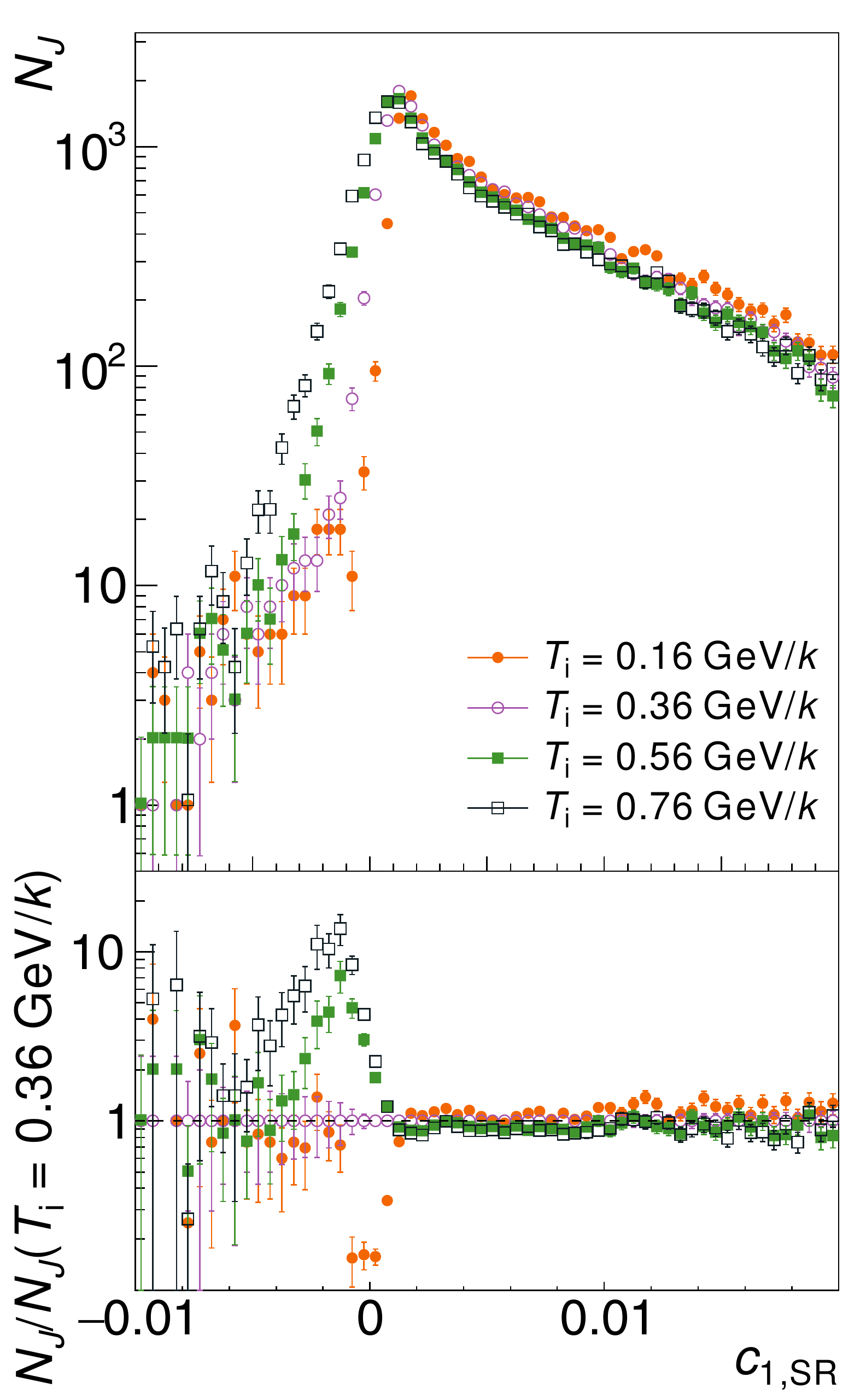}%
    \includegraphics[height=3.25in]{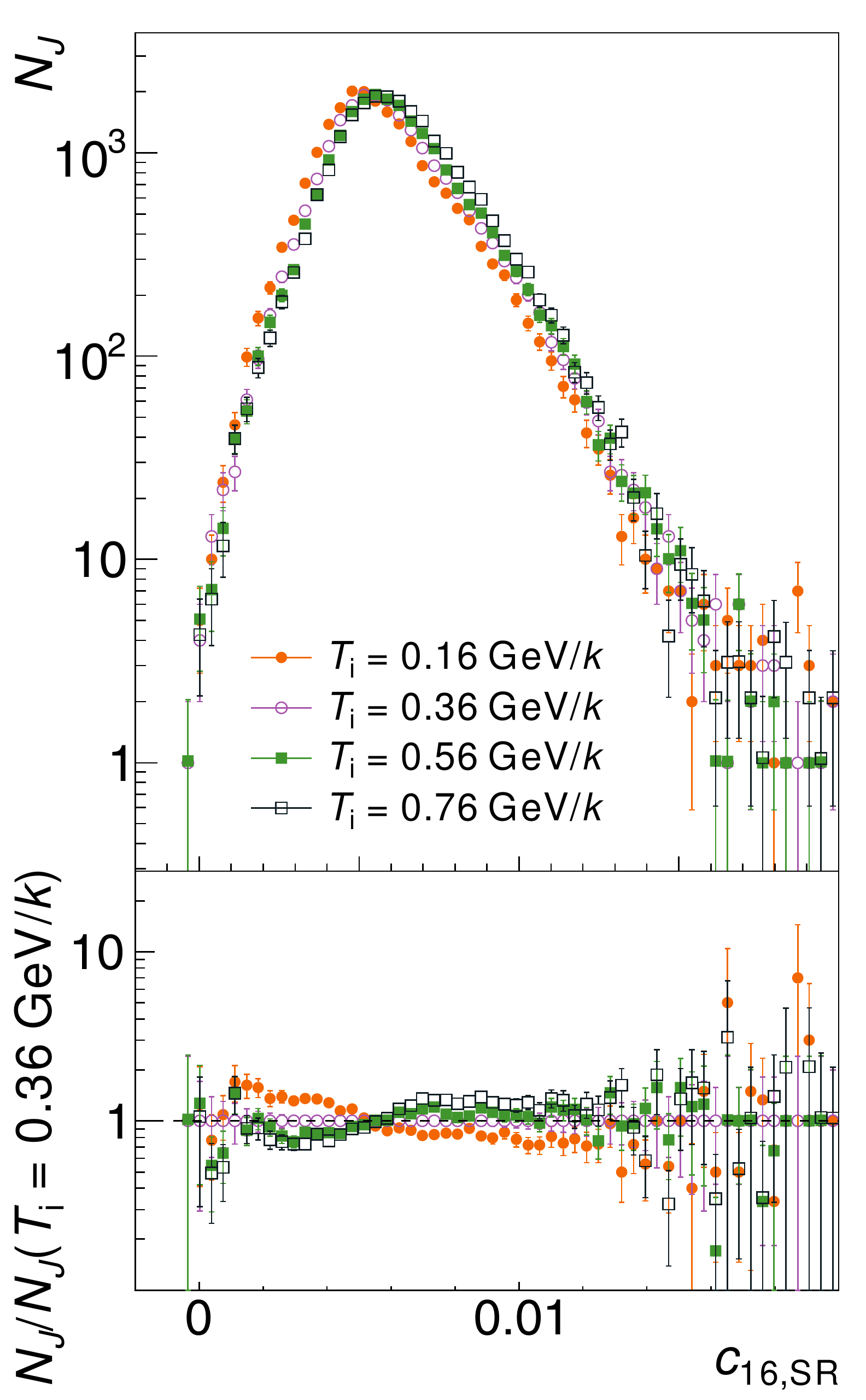}%
    \includegraphics[height=3.25in]{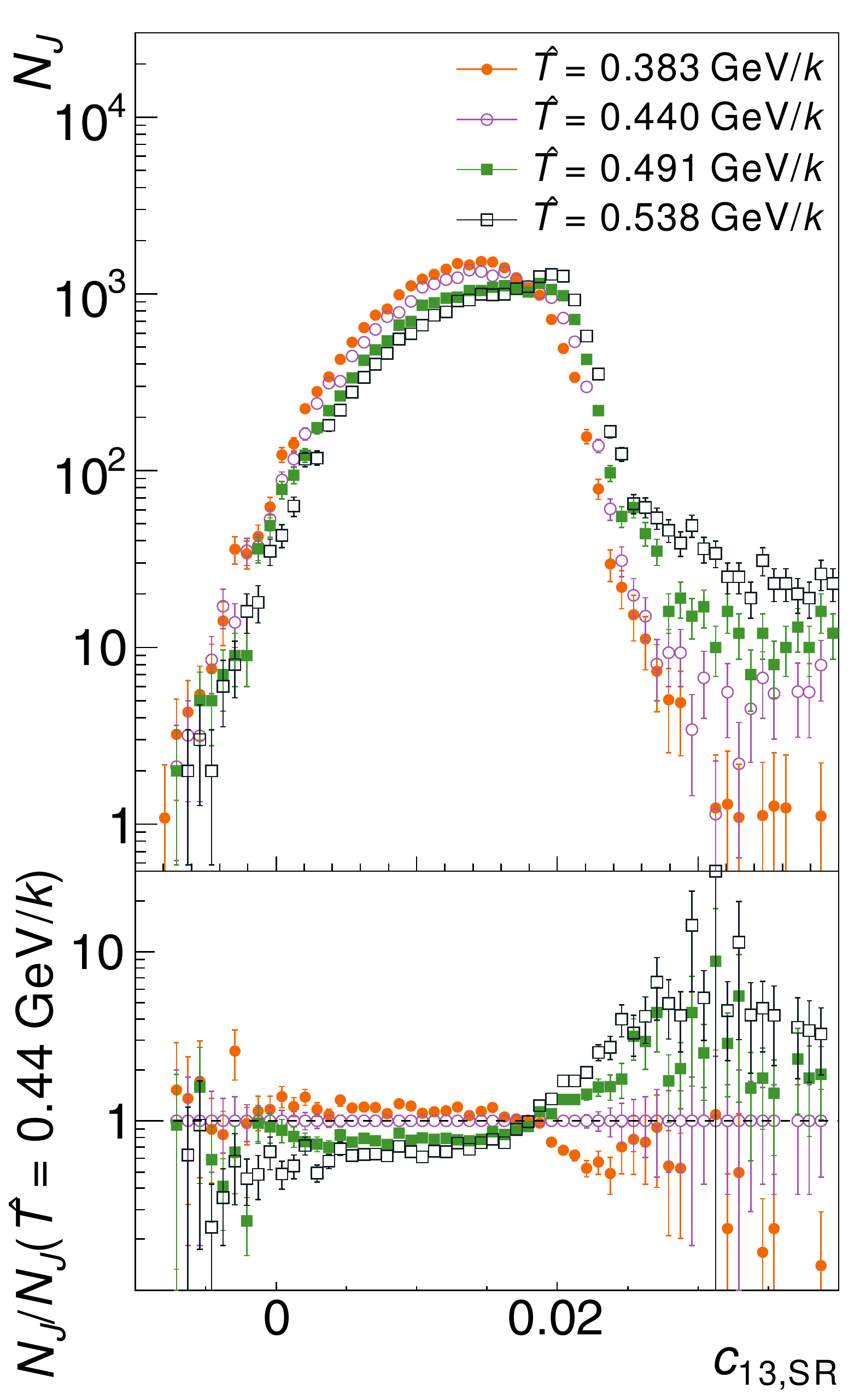}}
  \centerline{\hfill (a)\hfill\hfill (b)\hfill\hfill (c)\hfill}
  \caption{The distribution of the symbolic regression approximated
    neuron (a) $c_{1,\mathrm{SR}}$ and (b) $c_{16,\mathrm{SR}}$ for
    various $T_\mathrm{i}$ in \textsc{Jewel} for $100 < p_{T,J} <
    300\:\mathrm{GeV}/c$, with the ratio relative to $T_\mathrm{i} =
    0.36\:\mathrm{GeV}/k$. (c) shows the distribution of the symbolic
    regression approximated neuron $c_{13,\mathrm{SR}}$ for various
    $\hat{T}$ in the Linearized Boltzmann Transport model, with the
    ratio relative to $\hat{T} = 0.44\:\mathrm{GeV}/k$. In each case,
    $2\times 10^4$ jets are plotted}
  \label{fig:jewel_c1sr_c16sr}
\end{figure*}

For each jet, the energy flow polynomials (EFP)~\cite{Komiske2018},
i.e. intra-jet angular correlation of energies expressed as the
product of the relative momentum fraction carried by the final state
jet constituents and their relative opening angles, are calculated.
The EFP in this article is calculated for particles $p$ within a
$\Delta R_{pJ}^2 = (\eta_p - \eta_J)^2 + (\phi_p - \phi_J)^2 < D^2$
around the jet axis $J$, irrespective whether the particle has been
clustered into the jet by the jet reconstruction algorithm. This
choice accommodates purely calorimetric reconstruction of the jet
kinematics, whereas the substructure is then determined from tracking
detectors. Each EFP corresponds to a multigraph $G = (V, E)$, with the
vertices $V$ being $N_V$ particles that are inside the disk, and edges
$E$ a multiset consisting of pairs of vertices within $V$. The EFP of
$G$ is
\begin{equation}
  \begin{split}
    \mathrm{EFP}_G &= \sum_{j_1} \dots \sum_{j_{N_V}} \Biggl( z_{j_1}
    \dots z_{j_{N_V}} \prod_{(k, l) \in E} \theta_{kl} \Biggr)
  \end{split}
\end{equation}
where $z_j = p_{T,j}/p_{T,J}$, and $\theta_{kl}^2 = \Delta R_{kl}^2 =
(\eta_k - \eta_l)^2 + (\phi_k - \phi_l)^2$. The size of the multiset
$E$ is the degree of the polynomial. All $489$ primitive polynomials
with $1\le\text{degree}\le 7$ are used (note that the only polynomial
with degree 0 is the jet $p_{T,\mathrm{tot}}$), except for four
polynomials with an irreducible rank $(4, 4)$ tensor trace, as they
are prohibitively slow to calculate in the presence of $\approx 400$
particles inside a jet from central Pb-Pb collision ($\approx 6$
minutes per jet on an Intel Haswell Xeon at $3.5\:\mathrm{GHz}$). For
efficiency, evaluation of the remaining polynomials has been
reimplemented using
\textsc{blas}~\cite{doi:10.1177/10943420020160010101,*%
  2002:USB:567806.567807}. As the handling of UE by the Neural Network
is of interest, the jet substructure is not further manually corrected
to the hard scale.

As a second possible jet substructure variable, a polynomial expansion
of the jet shape in $\Delta R_{pJ}$, $(\eta_p - \eta_J)/(\phi_p -
\phi_J)$, and $z_p$ between the constituent particle $p$ and jet axis
$J$ was also explored. For an expansion into a comparable number of
coefficients, the performance was significantly below that of the EFP.

Fig.\ \ref{fig:nn_layout} shows the schematic layout of the NN
employed in this article. In this NN, $f(x) = d(s(c(x)))$ is a
function of the observable $x$, which is a series expansion of the
internal substructure formed by the final state particles of a
reconstructed jet. The function $f$ is composed of the following
groups of NN layers:
\begin{enumerate}
    \item The convolutional observable extraction layers
$c: \mathbb{R}^{K \times N} \mapsto \mathbb{R}^{M \times N}$, where
$N$ is the number of input jets, $K$ the number of input
features/$x$ observables per jet--event, and $M$ the number of
machine-learned observables per jet--event;
\item the symmetrization layer $s: \mathbb{R}^{M \times N} \mapsto
  \mathbb{R}^{M \times N}$;
\item the statistical analysis layers $d: \mathbb{R}^{M \times N}
  \mapsto \mathbb{R}$.
\end{enumerate}  

The presence of $s$ is needed, because the connection between neurons
in a NN are inherently sensitive to the ordering of its input. Placing
$s$ in front of $d$ however, allows one to ``retrofit'' an otherwise
ordered NN with permutation invariance. $M = 1$ would still satisfy
the universal approximation theorem, the presence of a small layer
creates a choke point inside the NN and potentially deteriorates the
convergence property observed in
\cite{pmlr-v38-choromanska15,*Haeffele2015,*Janzamin2015}.

In a NN, each layer $k$ is an operation of the form
\begin{equation}
  x_{k + 1}(x_k) = \sigma(W x_k + b)
  \label{eq:nn_layer}
\end{equation}
where the matrix $x_k$ is the input of the $k$-th layer, $W$ the
weight matrix, $b$ the bias vector, and $\sigma$ a nonlinear
activation function. Each neuron consists of the weight (a matrix
multiplied with the input vector), a bias (the vector added after the
matrix multiplication) and a suitable activation function. During the
course of training, the each neuron learns an "activation", which is a
real number usually close to $\pm 1$. Hence, a single neuron without
the activation function is a linear classifier. The activation
function -- termed analogously to the function of the firing rate in a
biological neuron -- is needed to produce nonlinear classification.

The initial layers $c$ represent per-jet--event analysis, which are
identical functions applied to a single jet--event, and repeated with
the $N$ jet--events. Each layer in $c$ is therefore a special case of
a one-dimensional, discrete convolution.

These convolutions of $N$ jet--events represent a sizeable statistical
sample, so the sample mean and standard deviation can be reliably
determined. A batch normalization~\cite{pmlr-v37-ioffe15} is made to
the $N$ jet--events in each layer in $c$, before application of the
activation function. In this step, the mean is subtracted from all
values; the differences are inversely scaled by the batch standard
deviations. This scales the values to be approximately in $[-1, 1]$,
and prevents neurons from becoming permanently ``stuck", i.e. neurons
which never activate across the entire training dataset. This is also
alleviated by making the network sufficiently large to allow for
redundancy.

The symmetrization operation $s$ transforms the $M\times N$ matrix
input $(x_{j\pi(k)})$, $j \in \{1, \ldots, M\}$, $k \in \{1, \ldots,
N\}$, into polynomials that enforce invariance under the identical
permutation $\pi$ of the columns corresponding to individual
jet--events, but breaks under a $(x_{j\pi_j(k)})$, if two rows exist
with different permutations $\pi_j$. This avoids limiting the NN to $M
= 1$, and differs from the construction using symmetric polynomials,
e.g.\ employed in~\cite{doi:10.1063/1.4817187}, which would describe a
function space other than that of possible statistical analyses. The
polynomials of order $m = 1$ are identical to the elementary symmetric
polynomials
\begin{equation}
  s_j(x) = \sum_{k = 1}^N x_{j\pi_j(k)}.
\end{equation}
A possible choice for $N = 2$ and $m = 2$ is
\begin{equation}
  \begin{split}
    s_1(x) &= x_{1\pi(1)} x_{2\pi(1)} + x_{1\pi(2)} x_{2\pi(2)}\\
    s_2(x) &= x_{1\pi(1)} x_{2\pi(2)} + x_{1\pi(2)} x_{2\pi(1)}
  \end{split}
\end{equation}
and for $N \ge 3$ and $m \ge 2$
\begin{equation}
  s_j(x) = \sum_{k = 1}^N x_{j\pi(k)}^{m - \lfloor m / 2\rfloor} x_{(j
    + 1)\mathop{\mathrm{mod}} M,\pi(k)}^l.
\end{equation}

Finally, $d$ consists of fully connected NN layers, where the final
layer outputs into two output neurons, one for each scenario of the
heavy-ion medium. The output neurons are trained with the goal of
being 1 for the correct medium scenario, and 0 for the incorrect one,
but will attain a probability-like value in between, when the
discrimination is imperfect.

The activation function has to be nonlinear (or the NN is reducible to
a purely linear model), and is chosen to be $\sigma(x_j) = \max(0,
x_j)$, the rectified linear unit (ReLU) that was found to be
efficiently trainable~\cite{Hahnloser2000,*pmlr-v15-glorot11a}. The
exception is the last layer, where the activation function is the
softmax function $\sigma(x_j) = \exp(x_j) / \sum_k \exp(x_k)$ that
converts a vector with multiple values into a probabilistic value
between 0 and 1~\cite{10.1007/978-3-642-76153-9_28}.

A NN trained without further constraints is difficult to analyze,
because the network's ability for discrimination is distributed among
all possible inputs, without an easy way to determine, \emph{a
  posteriori}, whether part of the NN is redundant or not contributing
to the overall performance. The NN employed here is regularized during
training using a metric $R$ that approximates how many input terms are
needed describe the function inside the NN, or the input complexity.
For a NN output after the $k$-th layer, the absolute partial
derivative from the $m$-th input to the $l$-th output is bounded by
\begin{equation}
  I_{lm} = \max \left\{ \left\lvert
  \inf\left(\frac{\partial x_{k,l}}{\partial x_{1,m}}\right)
  \right\rvert, \left\lvert \sup\left(\frac{\partial x_{k,l}}{\partial
    x_{1,m}}\right) \right\rvert \right\}
\end{equation}
where $\inf$, $\sup$ are the lower and upper interval arithmetic
bounds. The sum of absolute magnitude, or $\ell^1$ norm of $I_{lm}$,
inversely scaled by the maximum value
\begin{equation}
  R = \frac{\sum_{l,m} I_{lm}}{\max_{l,m} I_{lm}}.
  \label{eq:regularization}
\end{equation}
is then an upper bound of the input complexity that is easily
calculable during the NN training, and can be used to guide the NN off
configurations where an excessive number of input variables are used,
such that it becomes difficult to distinguish input essential
vs.\ redundant to the performance of the NN. Similar ideas exist in
the literature, like weight decay~\cite{NIPS1991_563} and the
layer-wise regularization by the Lipschitz
continuity~\cite{Cisse2017,*Gouk2018}. Unlike existing regularization
in the literature, $R$ targets the input complexity specifically, and
does not prevent hidden layers from using multiple neurons to form
nonlinear functions.

Fig.\ \ref{fig:jewel_regularization} shows how $I_{lm}$ for the layer
group $c$ evolves, when trained to discriminate $T_\mathrm{i}$ in
\textsc{Jewel}. The index $m \in \{1, 2, \ldots, 489\}$ counts the
energy flow polynomial with $1\le\text{degree}\le 7$, and $l$ is the
index of the intermediate analysis that $c$ outputs. The progression
is plotted after three different epochs, which are the number of full
passes through the training data. Arrows in the last training epoch
indicates the variable the symbolic regression used in
\eqref{eq:jewel_c1sr}--\eqref{eq:lbt_c13sr}, which are located in
regions of moderate value of $I_{lm}$. As $I_{lm}$ is only an upper
bound, it should not surprise that location relevant for the symbolic
regression are not the highest values of $I_{lm}$.

The 
function for minimization (loss function) by
the NN is 
\begin{equation}
  L = H + \mu_c R_c + \mu_d R_d
\end{equation}
where $H = -\log(p)$ the cross entropy, with $p$ being the probability
from multiple pseudoexperiments that the NN has determined the correct
medium scenario from $N$ jet--events. $R_c$ and $R_d$ are the values
of \eqref{eq:regularization} for layer groups $c$ and $d$. The
constants $\mu_c$, $\mu_d$ are regularization parameters that must be
adjusted to achieve a particular trade-off how many input terms the
trained NN will require to achieve its performance (being $H$),
vs.\ $H$ itself. This type of $\ell^1$-norm optimization with
competing objectives, that are linked together with an adjustable
parameter, is also referred to in the literature as the least absolute
shrinkage and selection operator (LASSO)
\cite{doi:10.1137/0907087,*10.2307/2346178}.

The NN is implemented in \textsc{TensorFlow}
1.10.0~\cite{tensorflow2015-whitepaper}, running on a Nvidia GP102 at
$1.4\:\mathrm{GHz}$ with \textsc{cuDNN} 7.1~\cite{ChetlurWVCTCS14}.
The symbolic regression (SR) algorithm
\textsc{ffx}~\cite{McConaghy2011,*6055329} is then applied to NN layer
groups $c$ and $d$ to extract closed-form expressions that
approximates the function of those NN layer groups.

\begin{figure}
  \includegraphics[width=3.375in]{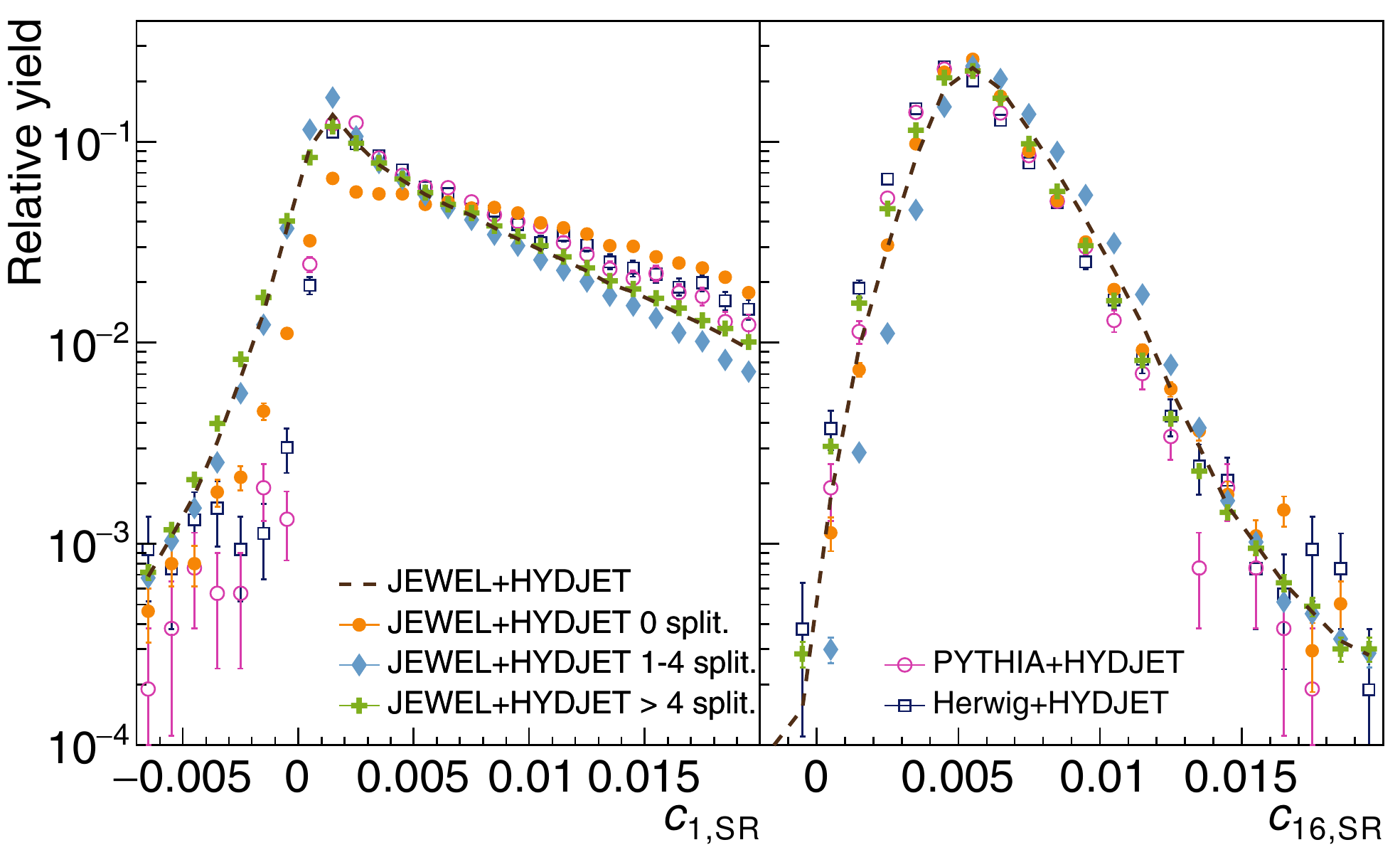}
  \caption{Dependence of the symbolic regression approximated
    $c_{1,\mathrm{SR}}$ and $c_{16,\mathrm{SR}}$ that discriminates
    $T_\mathrm{i}$ in \textsc{Jewel}, for various tagged number of
    splitting induced by jet--plasma interaction, normalized by the
    integral of each. The dashed line represents the distribution of
    the untagged \textsc{Jewel} (averaged over $T_\mathrm{i}$ and
    $\tau_\mathrm{i}$ scenarios). Also shown are the distributions
    from \textsc{pythia} 8.235 tune CUETP8M1 and Herwig 7.1.1 tune
    H7.1-Default, both embedded into the same Pb-Pb 0--10\% underlying
    event. Both the information on the splitting, and \textsc{pythia}
    8/Herwig 7 events, were not made available to the NN.}
  \label{fig:jewel_splitting}
\end{figure}

Training in the case of \textsc{Jewel} is performed on 65
pseudoexperiments, with each having a high and a low $T_\mathrm{i}$
(i.e.\ 130 pseudoexperiments in total). Each pseudoexperiment contains
2500 jet--events (not shared between pseudoexperiments). The Adam
optimization algorithm with its default parameters in \cite{KingmaB14}
is used. Whenever the optimization algorithm passes the entire
$130\times 2500$ unique events, the optimization is considered to have
completed an optimization epoch. After 1000 epochs using $\mu_c =
\mu_d = 0.01$, the accuracy is $0.973\pm 0.013$ (68\% Jeffreys
interval).

SR incrementally tests expressions with increasing size complexity,
and retains for each size complexity which form of expression achieved
the best accuracy. For \textsc{Jewel}, the overall analysis (extracted
using the first of the two output neurons) is approximated by
\begin{equation}
  d_{1,\mathrm{SR}} = 8.99 - 0.176 N \langle c_{16} \rangle - 0.175 N
  \langle c_1 \rangle
\end{equation}
where $\langle\rangle$ is the arithmetic mean, and $N = 2500$ is the
jet--event count per pseudoexperiment during the training. The
subscript indicates that the approximation $d_{1,\mathrm{SR}} \approx
d_1$ was made by the SR. This particular approximation is the highest
complex one that does not involve correlation between multiple jet
substructure observables, and has a normalized mean square error
(NMSE) of 10.9\%, vs. 28.3\% for a constant.

The activation
of $c_{1,\mathrm{SR}} \approx c_1$ and $c_{16,\mathrm{SR}} \approx
c_{16}$ are given by
\begin{equation}
\begin{split}
  c_{1,\mathrm{SR}} &= 0.0112 +\\
  &\quad + 45.5 \theta_{ab} \theta_{ac} \theta_{bc} \theta_{ad} \theta_{bd} \theta_{ae} \theta_{ce} z_a \cdots z_e +\\
  &\quad + 20.9 \theta_{ab} \theta_{ac} \theta_{bc} \theta_{ad} \theta_{bd} \theta_{ae} \theta_{be} z_a \cdots z_e +\\
  &\quad + 17.7 \theta_{ab} \theta_{ac} \theta_{bd} \theta_{cd} \theta_{ae} \theta_{de} z_a \cdots z_e +\\
  &\quad + 8.63 \theta_{ab} \theta_{ac} \theta_{bd} \theta_{cd} \theta_{ae} \theta_{de}^2 z_a \cdots z_e -\\
  &\quad - 3.08 \theta_{ab} \theta_{ac} \theta_{ad} \theta_{ae} z_a \cdots z_e -\\
  &\quad - 1.08 \theta_{ab} \theta_{ac} \theta_{ad}^2 \theta_{ae} z_a \cdots z_e +\\
  &\quad + 0.769 \theta_{ab} \theta_{ac} \theta_{ad} \theta_{ae} \theta_{af} \theta_{ag} \theta_{ah} z_a \cdots z_h -\\
  &\quad - 0.233 \theta_{ab}^3 \theta_{ac} \theta_{ad} \theta_{ae} z_a \cdots z_e -\\
  &\quad - 0.0377 \theta_{ab} \theta_{ac}^2 \theta_{ad} \theta_{ae}^2 z_a \cdots z_e -\\
  &\quad - 0.00483 \theta_{ab} \theta_{ac} \theta_{ad} \theta_{ae}^4 z_a \cdots z_e -\\
  &\quad - 0.000508 \theta_{ab} \theta_{ac}^3 \theta_{ad}^2 \theta_{ae} z_a \cdots z_e -\\
  &\quad - 4.51\times 10^{-5} \theta_{ab}^2 \theta_{ac} \theta_{ad}^2 \theta_{ae}^2 z_a \cdots z_e
\end{split}
\label{eq:jewel_c1sr}
\end{equation}
\begin{equation}
\begin{split}
  c_{16,\mathrm{SR}} &= 0.0362 +\\
  &\quad + 0.594 \theta_{ab}^3 \theta_{ac} \theta_{ad} \theta_{ae} z_a \cdots z_e +\\
  &\quad + 0.575 \theta_{ab} \theta_{ac}^2 \theta_{ad} \theta_{ae}^2 z_a \cdots z_e +\\
  &\quad + 0.421 \theta_{ab} \theta_{ac} \theta_{ad}^2 \theta_{ae} z_a \cdots z_e +\\
  &\quad + 0.420 \theta_{ab} \theta_{ac} \theta_{ad} \theta_{ae}^4 z_a \cdots z_e +\\
  &\quad + 0.246 \theta_{ab} \theta_{ac}^3 \theta_{ad}^2 \theta_{ae} z_a \cdots z_e +\\
  &\quad + 0.187 \theta_{ab} \theta_{ac} \theta_{ad} \theta_{ae} z_a \cdots z_e +\\
  &\quad + 0.120 \theta_{ab}^2 \theta_{ac} \theta_{ad}^2 \theta_{ae}^2 z_a \cdots z_e -\\
  &\quad - 0.0465 \theta_{ab}^3 z_a z_b - 0.0453 \theta_{ab}^4 z_a z_b -\\
  &\quad - 0.0333 \theta_{ab}^5 z_a z_b - 0.0328 \theta_{ab}^2 z_a z_b -\\
  &\quad - 0.0196 \theta_{ab}^6 z_a z_b - 0.0146 \theta_{ab} z_a z_b -\\
  &\quad - 0.00963 \theta_{ab}^7 z_a z_b
\end{split}
\label{eq:jewel_c16sr}
\end{equation}
where the Einstein summation is implied over $a, b, \ldots, h$, $z_a
z_b z_c z_d z_e$ has been shortened into $z_a \cdots z_e$, and $z_a
\cdots z_e z_f z_g z_h$ into $z_a \cdots z_h$.

For LBT, discrimination is far easier to achieve, and the sample size
can be reduced to 50 pseudoexperiments (100 for both $\hat{T}$
scenarios) containing $N = 1200$ jet--events each. Setting the
regularization at $\mu_1 = \mu_2 = 30$ was able to yield a validation
accuracy of $0.975\pm 0.014$, and the corresponding SR approximated
analysis
\begin{equation}
  d_{1,\mathrm{SR}} = 4.49 - 0.318 N \langle c_{13}\rangle - 0.00653 N
  \langle c_{13}\rangle^2
\end{equation}
(NMSE is 1.17\%, vs. 36.9\% for a constant)
involves a single variable, which is approximated by SR as
\begin{equation}
\begin{split}
  c_{13,\mathrm{SR}} &= 0.0453 -\\
  &\quad - 0.00109 \log_{10}(p_1) (\log_{10}(p_2) + \log_{10}(p_3)) -\\
  &\quad - 0.000829 \log_{10}(p_2) \log_{10}(p_3)
\end{split}
\label{eq:lbt_c13sr}
\end{equation}
where
\begin{equation}
\begin{split}
p_1 &= \theta_{ab} \theta_{ac} \theta_{bc} \theta_{ad} \theta_{bd} \theta_{ae} \theta_{be} z_a \cdots z_e\\
p_2 &= \theta_{ab} \theta_{ac} \theta_{bd} \theta_{cd} \theta_{ae} \theta_{de} z_a \cdots z_e\\
p_3 &= \theta_{ab} \theta_{ac} \theta_{bd} \theta_{cd} \theta_{ae} \theta_{de}^2 z_a \cdots z_e
\end{split}
\end{equation}
The $\log_{10}$ function appears to be selected by SR as a function of
convenience to represent the range compression inside the NN, and is
unlikely to have deeper meaning.

Fig.\ \ref{fig:jewel_c1sr_c16sr} shows the distribution of
$c_{1,\mathrm{SR}}$ and $c_{16,\mathrm{SR}}$ for various
\textsc{Jewel} $T_\mathrm{i}$, and $c_{13,\mathrm{SR}}$ for LBT and
various $\hat{T}$. The $c_{1,\mathrm{SR}} < 0$ for \textsc{Jewel} is a
particularly striking region, where $T_\mathrm{i}$ induces a change by
an order of magnitude.

From the form of the expression obtained, one can see that angular
correlations between five particles are frequently used. To
investigate this further, \textsc{Jewel} was modified to allow
book-keeping of the angular direction of the originating parton each
time a splitting occurred. For each reconstructed jet, the number of
splittings within the angular extent of the jet is used to count
splittings due to plasma medium interaction. Additionally, the Lund
string model based \textsc{pythia} 8.235 tune CUETP8M1, and Herwig
7.1.1 tune H7.1-Default~\cite{Baehr2008,*Bellm2016} that is based on
the Webber model of cluster fragmentation, both embedded into the
identical Pb-Pb 0--10\% UE as for \textsc{Jewel} and LBT, were
included to check for the p-p expectation including potential
fragmentation model dependence.

Fig.\ \ref{fig:jewel_splitting} shows the distribution of
$c_{1,\mathrm{SR}}$ and $c_{16,\mathrm{SR}}$ in \textsc{Jewel} with 0,
1--4, and $> 4$ splittings. Overlaid are also \textsc{pythia} 8 and
Herwig 7 embedded into Pb-Pb. The two variables exhibit significant
dependence on the number of plasma-induced splittings, with no
significant fragmentation function dependence upon comparing
\textsc{pythia} 8 to Herwig 7. The variable $c_{16,\mathrm{SR}}$
evolves with $T_\mathrm{i}$ by tagging jets with 1--4 splittings,
while losing resolving power thereafter ($c_{16,\mathrm{SR}}$ for
those jets resemble the average p-p jet). On the other hand,
$c_{1,\mathrm{SR}}$ has a region $c_{1,\mathrm{SR}} < 0$ that is
additionally sensitive to $> 4$ splittings.

Interestingly, the same $c_{1,\mathrm{SR}} < 0$ (a quantile containing
$6.65\pm 0.04\%$ of the jets in \textsc{Jewel} averaged over the
$T_\mathrm{i}$ and $\tau_\mathrm{i}$ scenarios) is populated due to
the recoiling scattering centers in \textsc{Jewel}. Its significance
vs. \textsc{pythia} 8 ($0.68\pm 0.11\%$ of the jets) and Herwig 7
($1.4\pm 0.2\%$ of the jets) provides a strong constraint whether the
recoil effect as suggested by \textsc{Jewel} exists. The only known
indication of a similar effect between \textsc{Jewel} recoil on and
off is the groomed jet mass, which was recently studied by
CMS~\cite{Sirunyan:2018gct}. One should point out the NN learned of
the \textsc{Jewel} recoil incidentally, and not by a training
procedure targeting this effect. And unlike human experts, the NN was
never able to observe \textsc{Jewel} with either the recoil effects
switched off, or without the full Pb-Pb UE. This hints that the
machine learning technique presented here, operating on a commodity,
off-the-self hardware, is at least able to study the characteristics
of \textsc{Jewel} to a detail comparable to the human experts in the
current heavy-ion field.

This article describes a system of machine learning for the discovery
of jet substructure analyses, using a NN structured to automatically
learn statistical analyses. The approach includes regularization to
sufficiently apply a simplification process during the training.
Symbolic Regression is then used to extract properties of the
resulting NN, and can reduce to compact, closed-form expressions
readily repeatable by humans. Applied to \textsc{Jewel} and LBT, two
heavy-ion jet MC event generators, it constructs analyses that can
reliably extract the initial temperature, at the presence of full bulk
Underlying Event -- a performance not previously demonstrated by human
constructed jet substructure analyses. In \textsc{Jewel}, the type of
jet substructure observables being constructed by machine learning is
found to be strongly dependent on, i.e. tagging, the number of
splittings experienced by a parton transiting a quark gluon plasma
medium. One of the discovered observables also tags the medium recoil
from the interaction, which is otherwise only known to be measurable
via the groomed jet mass. This indicates that the machine learning
technique described here performs original research with jet
substructure observables, at a level comparable to the human experts
in the field.

I would like to thank Barbara Jacak for discussions and critical
reading, additionally Xin-Nian Wang and Yayun He for the Linearized
Boltzmann Transport model and insightful suggestions.

\bibliographystyle{apsrev4-1}
\bibliography{cs,experiment,flow,generator,jet,math}

\begin{thebibliography}{58}%
\makeatletter
\providecommand \@ifxundefined [1]{%
 \@ifx{#1\undefined}
}%
\providecommand \@ifnum [1]{%
 \ifnum #1\expandafter \@firstoftwo
 \else \expandafter \@secondoftwo
 \fi
}%
\providecommand \@ifx [1]{%
 \ifx #1\expandafter \@firstoftwo
 \else \expandafter \@secondoftwo
 \fi
}%
\providecommand \natexlab [1]{#1}%
\providecommand \enquote  [1]{``#1''}%
\providecommand \bibnamefont  [1]{#1}%
\providecommand \bibfnamefont [1]{#1}%
\providecommand \citenamefont [1]{#1}%
\providecommand \href@noop [0]{\@secondoftwo}%
\providecommand \href [0]{\begingroup \@sanitize@url \@href}%
\providecommand \@href[1]{\@@startlink{#1}\@@href}%
\providecommand \@@href[1]{\endgroup#1\@@endlink}%
\providecommand \@sanitize@url [0]{\catcode `\\12\catcode `\$12\catcode
  `\&12\catcode `\#12\catcode `\^12\catcode `\_12\catcode `\%12\relax}%
\providecommand \@@startlink[1]{}%
\providecommand \@@endlink[0]{}%
\providecommand \url  [0]{\begingroup\@sanitize@url \@url }%
\providecommand \@url [1]{\endgroup\@href {#1}{\urlprefix }}%
\providecommand \urlprefix  [0]{URL }%
\providecommand \Eprint [0]{\href }%
\providecommand \doibase [0]{http://dx.doi.org/}%
\providecommand \selectlanguage [0]{\@gobble}%
\providecommand \bibinfo  [0]{\@secondoftwo}%
\providecommand \bibfield  [0]{\@secondoftwo}%
\providecommand \translation [1]{[#1]}%
\providecommand \BibitemOpen [0]{}%
\providecommand \bibitemStop [0]{}%
\providecommand \bibitemNoStop [0]{.\EOS\space}%
\providecommand \EOS [0]{\spacefactor3000\relax}%
\providecommand \BibitemShut  [1]{\csname bibitem#1\endcsname}%
\let\auto@bib@innerbib\@empty
\bibitem [{\citenamefont {Heinz}\ and\ \citenamefont
  {Snellings}(2013)}]{doi:10.1146/annurev-nucl-102212-170540}%
  \BibitemOpen
  \bibfield  {author} {\bibinfo {author} {\bibfnamefont {U.}~\bibnamefont
  {Heinz}}\ and\ \bibinfo {author} {\bibfnamefont {R.}~\bibnamefont
  {Snellings}},\ }\href {\doibase 10.1146/annurev-nucl-102212-170540}
  {\bibfield  {journal} {\bibinfo  {journal} {Ann. Rev. Nucl. Part. Sci.}\
  }\textbf {\bibinfo {volume} {63}},\ \bibinfo {pages} {123} (\bibinfo {year}
  {2013})}\BibitemShut {NoStop}%
\bibitem [{\citenamefont {Adcox}\ \emph {et~al.}(2005)\citenamefont {Adcox}
  \emph {et~al.}}]{ADCOX2005184}%
  \BibitemOpen
  \bibfield  {author} {\bibinfo {author} {\bibfnamefont {K.}~\bibnamefont
  {Adcox}} \emph {et~al.},\ }\href {\doibase 10.1016/j.nuclphysa.2005.03.086}
  {\bibfield  {journal} {\bibinfo  {journal} {Nucl. Phys. A}\ }\textbf
  {\bibinfo {volume} {757}},\ \bibinfo {pages} {184} (\bibinfo {year}
  {2005})}\BibitemShut {NoStop}%
\bibitem [{\citenamefont {Adams}\ \emph {et~al.}(2005)\citenamefont {Adams}
  \emph {et~al.}}]{ADAMS2005102}%
  \BibitemOpen
  \bibfield  {author} {\bibinfo {author} {\bibfnamefont {J.}~\bibnamefont
  {Adams}} \emph {et~al.},\ }\href {\doibase 10.1016/j.nuclphysa.2005.03.085}
  {\bibfield  {journal} {\bibinfo  {journal} {Nucl. Phys. A}\ }\textbf
  {\bibinfo {volume} {757}},\ \bibinfo {pages} {102} (\bibinfo {year}
  {2005})}\BibitemShut {NoStop}%
\bibitem [{\citenamefont {Mehtar-Tani}\ and\ \citenamefont
  {Tywoniuk}(2017)}]{Mehtar-Tani2017}%
  \BibitemOpen
  \bibfield  {author} {\bibinfo {author} {\bibfnamefont {Y.}~\bibnamefont
  {Mehtar-Tani}}\ and\ \bibinfo {author} {\bibfnamefont {K.}~\bibnamefont
  {Tywoniuk}},\ }\href {\doibase 10.1007/JHEP04(2017)125} {\bibfield  {journal}
  {\bibinfo  {journal} {Journal of High Energy Physics}\ }\textbf {\bibinfo
  {volume} {2017}},\ \bibinfo {pages} {125} (\bibinfo {year}
  {2017})}\BibitemShut {NoStop}%
\bibitem [{\citenamefont {Chien}\ and\ \citenamefont
  {Vitev}(2017)}]{PhysRevLett.119.112301}%
  \BibitemOpen
  \bibfield  {author} {\bibinfo {author} {\bibfnamefont {Y.-T.}\ \bibnamefont
  {Chien}}\ and\ \bibinfo {author} {\bibfnamefont {I.}~\bibnamefont {Vitev}},\
  }\href {\doibase 10.1103/PhysRevLett.119.112301} {\bibfield  {journal}
  {\bibinfo  {journal} {Phys. Rev. Lett.}\ }\textbf {\bibinfo {volume} {119}},\
  \bibinfo {pages} {112301} (\bibinfo {year} {2017})}\BibitemShut {NoStop}%
\bibitem [{\citenamefont {Kunnawalkam~Elayavalli}\ and\ \citenamefont
  {Zapp}(2017)}]{KunnawalkamElayavalli2017}%
  \BibitemOpen
  \bibfield  {author} {\bibinfo {author} {\bibfnamefont {R.}~\bibnamefont
  {Kunnawalkam~Elayavalli}}\ and\ \bibinfo {author} {\bibfnamefont {K.~C.}\
  \bibnamefont {Zapp}},\ }\href {\doibase 10.1007/JHEP07(2017)141} {\bibfield
  {journal} {\bibinfo  {journal} {Journal of High Energy Physics}\ }\textbf
  {\bibinfo {volume} {2017}},\ \bibinfo {pages} {141} (\bibinfo {year}
  {2017})}\BibitemShut {NoStop}%
\bibitem [{\citenamefont {Sirunyan}\ \emph
  {et~al.}(2018{\natexlab{a}})\citenamefont {Sirunyan} \emph
  {et~al.}}]{PhysRevLett.120.142302}%
  \BibitemOpen
  \bibfield  {author} {\bibinfo {author} {\bibfnamefont {A.~M.}\ \bibnamefont
  {Sirunyan}} \emph {et~al.} (\bibinfo {collaboration} {CMS Collaboration}),\
  }\href {\doibase 10.1103/PhysRevLett.120.142302} {\bibfield  {journal}
  {\bibinfo  {journal} {Phys. Rev. Lett.}\ }\textbf {\bibinfo {volume} {120}},\
  \bibinfo {pages} {142302} (\bibinfo {year} {2018}{\natexlab{a}})}\BibitemShut
  {NoStop}%
\bibitem [{\citenamefont {Song}\ \emph
  {et~al.}(2011{\natexlab{a}})\citenamefont {Song}, \citenamefont {Bass},
  \citenamefont {Heinz}, \citenamefont {Hirano},\ and\ \citenamefont
  {Shen}}]{PhysRevLett.106.192301}%
  \BibitemOpen
  \bibfield  {author} {\bibinfo {author} {\bibfnamefont {H.}~\bibnamefont
  {Song}}, \bibinfo {author} {\bibfnamefont {S.~A.}\ \bibnamefont {Bass}},
  \bibinfo {author} {\bibfnamefont {U.}~\bibnamefont {Heinz}}, \bibinfo
  {author} {\bibfnamefont {T.}~\bibnamefont {Hirano}}, \ and\ \bibinfo {author}
  {\bibfnamefont {C.}~\bibnamefont {Shen}},\ }\href {\doibase
  10.1103/PhysRevLett.106.192301} {\bibfield  {journal} {\bibinfo  {journal}
  {Phys. Rev. Lett.}\ }\textbf {\bibinfo {volume} {106}},\ \bibinfo {pages}
  {192301} (\bibinfo {year} {2011}{\natexlab{a}})}\BibitemShut {NoStop}%
\bibitem [{\citenamefont {Luzum}(2011)}]{PhysRevC.83.044911}%
  \BibitemOpen
  \bibfield  {author} {\bibinfo {author} {\bibfnamefont {M.}~\bibnamefont
  {Luzum}},\ }\href {\doibase 10.1103/PhysRevC.83.044911} {\bibfield  {journal}
  {\bibinfo  {journal} {Phys. Rev. C}\ }\textbf {\bibinfo {volume} {83}},\
  \bibinfo {pages} {044911} (\bibinfo {year} {2011})}\BibitemShut {NoStop}%
\bibitem [{\citenamefont {Song}\ \emph
  {et~al.}(2011{\natexlab{b}})\citenamefont {Song}, \citenamefont {Bass},\ and\
  \citenamefont {Heinz}}]{PhysRevC.83.054912}%
  \BibitemOpen
  \bibfield  {author} {\bibinfo {author} {\bibfnamefont {H.}~\bibnamefont
  {Song}}, \bibinfo {author} {\bibfnamefont {S.~A.}\ \bibnamefont {Bass}}, \
  and\ \bibinfo {author} {\bibfnamefont {U.}~\bibnamefont {Heinz}},\ }\href
  {\doibase 10.1103/PhysRevC.83.054912} {\bibfield  {journal} {\bibinfo
  {journal} {Phys. Rev. C}\ }\textbf {\bibinfo {volume} {83}},\ \bibinfo
  {pages} {054912} (\bibinfo {year} {2011}{\natexlab{b}})}\BibitemShut
  {NoStop}%
\bibitem [{\citenamefont {Gale}\ \emph {et~al.}(2013)\citenamefont {Gale},
  \citenamefont {Jeon}, \citenamefont {Schenke}, \citenamefont {Tribedy},\ and\
  \citenamefont {Venugopalan}}]{PhysRevLett.110.012302}%
  \BibitemOpen
  \bibfield  {author} {\bibinfo {author} {\bibfnamefont {C.}~\bibnamefont
  {Gale}}, \bibinfo {author} {\bibfnamefont {S.}~\bibnamefont {Jeon}}, \bibinfo
  {author} {\bibfnamefont {B.}~\bibnamefont {Schenke}}, \bibinfo {author}
  {\bibfnamefont {P.}~\bibnamefont {Tribedy}}, \ and\ \bibinfo {author}
  {\bibfnamefont {R.}~\bibnamefont {Venugopalan}},\ }\href {\doibase
  10.1103/PhysRevLett.110.012302} {\bibfield  {journal} {\bibinfo  {journal}
  {Phys. Rev. Lett.}\ }\textbf {\bibinfo {volume} {110}},\ \bibinfo {pages}
  {012302} (\bibinfo {year} {2013})}\BibitemShut {NoStop}%
\bibitem [{\citenamefont {Hornik}\ \emph {et~al.}(1989)\citenamefont {Hornik},
  \citenamefont {Stinchcombe},\ and\ \citenamefont {White}}]{HORNIK1989359}%
  \BibitemOpen
  \bibfield  {author} {\bibinfo {author} {\bibfnamefont {K.}~\bibnamefont
  {Hornik}}, \bibinfo {author} {\bibfnamefont {M.}~\bibnamefont {Stinchcombe}},
  \ and\ \bibinfo {author} {\bibfnamefont {H.}~\bibnamefont {White}},\ }\href
  {\doibase 10.1016/0893-6080(89)90020-8} {\bibfield  {journal} {\bibinfo
  {journal} {Neural Networks}\ }\textbf {\bibinfo {volume} {2}},\ \bibinfo
  {pages} {359} (\bibinfo {year} {1989})}\BibitemShut {NoStop}%
\bibitem [{\citenamefont {Leshno}\ \emph {et~al.}(1993)\citenamefont {Leshno},
  \citenamefont {Lin}, \citenamefont {Pinkus},\ and\ \citenamefont
  {Schocken}}]{LESHNO1993861}%
  \BibitemOpen
  \bibfield  {author} {\bibinfo {author} {\bibfnamefont {M.}~\bibnamefont
  {Leshno}}, \bibinfo {author} {\bibfnamefont {V.~Y.}\ \bibnamefont {Lin}},
  \bibinfo {author} {\bibfnamefont {A.}~\bibnamefont {Pinkus}}, \ and\ \bibinfo
  {author} {\bibfnamefont {S.}~\bibnamefont {Schocken}},\ }\href {\doibase
  https://doi.org/10.1016/S0893-6080(05)80131-5} {\bibfield  {journal}
  {\bibinfo  {journal} {Neural Networks}\ }\textbf {\bibinfo {volume} {6}},\
  \bibinfo {pages} {861} (\bibinfo {year} {1993})}\BibitemShut {NoStop}%
\bibitem [{\citenamefont {Mhaskar}\ and\ \citenamefont
  {Poggio}(2016)}]{MhaskarP16}%
  \BibitemOpen
  \bibfield  {author} {\bibinfo {author} {\bibfnamefont {H.}~\bibnamefont
  {Mhaskar}}\ and\ \bibinfo {author} {\bibfnamefont {T.~A.}\ \bibnamefont
  {Poggio}},\ }\href@noop {} {\enquote {\bibinfo {title} {Deep vs.\ shallow
  networks: An approximation theory perspective},}\ } (\bibinfo {year}
  {2016}),\ \Eprint {http://arxiv.org/abs/1608.03287} {arXiv:1608.03287
  [cs.LG]} \BibitemShut {NoStop}%
\bibitem [{\citenamefont {Liang}\ and\ \citenamefont
  {Srikant}(2016)}]{LiangS16}%
  \BibitemOpen
  \bibfield  {author} {\bibinfo {author} {\bibfnamefont {S.}~\bibnamefont
  {Liang}}\ and\ \bibinfo {author} {\bibfnamefont {R.}~\bibnamefont
  {Srikant}},\ }\href@noop {} {\enquote {\bibinfo {title} {Why deep neural
  networks for function approximation?}}\ } (\bibinfo {year} {2016}),\ \Eprint
  {http://arxiv.org/abs/1610.04161} {arXiv:1610.04161 [cs.LG]} \BibitemShut
  {NoStop}%
\bibitem [{\citenamefont {Yarotsky}(2017)}]{YAROTSKY2017103}%
  \BibitemOpen
  \bibfield  {author} {\bibinfo {author} {\bibfnamefont {D.}~\bibnamefont
  {Yarotsky}},\ }\href {\doibase 10.1016/j.neunet.2017.07.002} {\bibfield
  {journal} {\bibinfo  {journal} {Neural Netw.}\ }\textbf {\bibinfo {volume}
  {94}},\ \bibinfo {pages} {103} (\bibinfo {year} {2017})}\BibitemShut
  {NoStop}%
\bibitem [{\citenamefont {\.{Z}ytkow}\ \emph {et~al.}(1990)\citenamefont
  {\.{Z}ytkow}, \citenamefont {Zhu},\ and\ \citenamefont
  {Hussam}}]{Zytkow:1990:ADC:1865609.1865633}%
  \BibitemOpen
  \bibfield  {author} {\bibinfo {author} {\bibfnamefont {J.~M.}\ \bibnamefont
  {\.{Z}ytkow}}, \bibinfo {author} {\bibfnamefont {J.}~\bibnamefont {Zhu}}, \
  and\ \bibinfo {author} {\bibfnamefont {A.}~\bibnamefont {Hussam}},\ }in\
  \href {https://www.aaai.org/Library/AAAI/1990/aaai90-133.php} {\emph
  {\bibinfo {booktitle} {Proceedings of the Eighth National Conference on
  Artificial Intelligence - Volume 2}}},\ \bibinfo {series and number}
  {AAAI'90}\ (\bibinfo  {publisher} {AAAI Press},\ \bibinfo {year} {1990})\
  pp.\ \bibinfo {pages} {889--894}\BibitemShut {NoStop}%
\bibitem [{\citenamefont {Lindsay}\ \emph {et~al.}(1993)\citenamefont
  {Lindsay}, \citenamefont {Buchanan}, \citenamefont {Feigenbaum},\ and\
  \citenamefont {Lederberg}}]{LINDSAY1993209}%
  \BibitemOpen
  \bibfield  {author} {\bibinfo {author} {\bibfnamefont {R.~K.}\ \bibnamefont
  {Lindsay}}, \bibinfo {author} {\bibfnamefont {B.~G.}\ \bibnamefont
  {Buchanan}}, \bibinfo {author} {\bibfnamefont {E.~A.}\ \bibnamefont
  {Feigenbaum}}, \ and\ \bibinfo {author} {\bibfnamefont {J.}~\bibnamefont
  {Lederberg}},\ }\href {\doibase 10.1016/0004-3702(93)90068-M} {\bibfield
  {journal} {\bibinfo  {journal} {Artificial Intelligence}\ }\textbf {\bibinfo
  {volume} {61}},\ \bibinfo {pages} {209 } (\bibinfo {year}
  {1993})}\BibitemShut {NoStop}%
\bibitem [{\citenamefont {King}\ \emph {et~al.}(2004)\citenamefont {King} \emph
  {et~al.}}]{King2004}%
  \BibitemOpen
  \bibfield  {author} {\bibinfo {author} {\bibfnamefont {R.~D.}\ \bibnamefont
  {King}} \emph {et~al.},\ }\href {\doibase 10.1038/nature02236} {\bibfield
  {journal} {\bibinfo  {journal} {Nature}\ }\textbf {\bibinfo {volume} {427}},\
  \bibinfo {pages} {247} (\bibinfo {year} {2004})}\BibitemShut {NoStop}%
\bibitem [{\citenamefont {Voytek}\ and\ \citenamefont
  {Voytek}(2012)}]{VOYTEK201292}%
  \BibitemOpen
  \bibfield  {author} {\bibinfo {author} {\bibfnamefont {J.~B.}\ \bibnamefont
  {Voytek}}\ and\ \bibinfo {author} {\bibfnamefont {B.}~\bibnamefont
  {Voytek}},\ }\href {\doibase https://doi.org/10.1016/j.jneumeth.2012.04.019}
  {\bibfield  {journal} {\bibinfo  {journal} {J. Neurosci. Methods}\ }\textbf
  {\bibinfo {volume} {208}},\ \bibinfo {pages} {92 } (\bibinfo {year}
  {2012})}\BibitemShut {NoStop}%
\bibitem [{\citenamefont {De~Vesine}\ \emph {et~al.}(2013)\citenamefont
  {De~Vesine}, \citenamefont {Anderson}, \citenamefont {Zreda}, \citenamefont
  {Zweck},\ and\ \citenamefont {Bradley}}]{72d570723c644a8cbb07011a2d39d526}%
  \BibitemOpen
  \bibfield  {author} {\bibinfo {author} {\bibfnamefont {L.~R.}\ \bibnamefont
  {De~Vesine}}, \bibinfo {author} {\bibfnamefont {K.}~\bibnamefont {Anderson}},
  \bibinfo {author} {\bibfnamefont {M.}~\bibnamefont {Zreda}}, \bibinfo
  {author} {\bibfnamefont {C.}~\bibnamefont {Zweck}}, \ and\ \bibinfo {author}
  {\bibfnamefont {L.}~\bibnamefont {Bradley}},\ }in\ \href
  {https://www.aaai.org/ocs/index.php/FSS/FSS13/paper/view/7556} {\emph
  {\bibinfo {booktitle} {Discovery Informatics: AI Takes a Science-Centered
  View on Big Data}}},\ \bibinfo {series} {AAAI Fall Symposium -- Technical
  Report}, Vol.\ \bibinfo {volume} {FS-13-01}\ (\bibinfo  {publisher} {AI
  Access Foundation},\ \bibinfo {year} {2013})\ pp.\ \bibinfo {pages}
  {16--22}\BibitemShut {NoStop}%
\bibitem [{\citenamefont {Spangler}\ \emph {et~al.}(2014)\citenamefont
  {Spangler} \emph {et~al.}}]{Spangler:2014:AHG:2623330.2623667}%
  \BibitemOpen
  \bibfield  {author} {\bibinfo {author} {\bibfnamefont {S.}~\bibnamefont
  {Spangler}} \emph {et~al.},\ }in\ \href {\doibase 10.1145/2623330.2623667}
  {\emph {\bibinfo {booktitle} {Proceedings of the 20th ACM SIGKDD
  International Conference on Knowledge Discovery and Data Mining}}},\ \bibinfo
  {series and number} {KDD '14}\ (\bibinfo  {publisher} {ACM},\ \bibinfo
  {address} {New York, NY, USA},\ \bibinfo {year} {2014})\ pp.\ \bibinfo
  {pages} {1877--1886}\BibitemShut {NoStop}%
\bibitem [{\citenamefont {Kadurin}\ \emph {et~al.}(2017)\citenamefont
  {Kadurin}, \citenamefont {Nikolenko}, \citenamefont {Khrabrov}, \citenamefont
  {Aliper},\ and\ \citenamefont
  {Zhavoronkov}}]{doi:10.1021/acs.molpharmaceut.7b00346}%
  \BibitemOpen
  \bibfield  {author} {\bibinfo {author} {\bibfnamefont {A.}~\bibnamefont
  {Kadurin}}, \bibinfo {author} {\bibfnamefont {S.}~\bibnamefont {Nikolenko}},
  \bibinfo {author} {\bibfnamefont {K.}~\bibnamefont {Khrabrov}}, \bibinfo
  {author} {\bibfnamefont {A.}~\bibnamefont {Aliper}}, \ and\ \bibinfo {author}
  {\bibfnamefont {A.}~\bibnamefont {Zhavoronkov}},\ }\href {\doibase
  10.1021/acs.molpharmaceut.7b00346} {\bibfield  {journal} {\bibinfo  {journal}
  {Mol. Pharm.}\ }\textbf {\bibinfo {volume} {14}},\ \bibinfo {pages} {3098}
  (\bibinfo {year} {2017})}\BibitemShut {NoStop}%
\bibitem [{\citenamefont {Krenn}\ \emph {et~al.}(2016)\citenamefont {Krenn},
  \citenamefont {Malik}, \citenamefont {Fickler}, \citenamefont {Lapkiewicz},\
  and\ \citenamefont {Zeilinger}}]{PhysRevLett.116.090405}%
  \BibitemOpen
  \bibfield  {author} {\bibinfo {author} {\bibfnamefont {M.}~\bibnamefont
  {Krenn}}, \bibinfo {author} {\bibfnamefont {M.}~\bibnamefont {Malik}},
  \bibinfo {author} {\bibfnamefont {R.}~\bibnamefont {Fickler}}, \bibinfo
  {author} {\bibfnamefont {R.}~\bibnamefont {Lapkiewicz}}, \ and\ \bibinfo
  {author} {\bibfnamefont {A.}~\bibnamefont {Zeilinger}},\ }\href {\doibase
  10.1103/PhysRevLett.116.090405} {\bibfield  {journal} {\bibinfo  {journal}
  {Phys. Rev. Lett.}\ }\textbf {\bibinfo {volume} {116}},\ \bibinfo {pages}
  {090405} (\bibinfo {year} {2016})}\BibitemShut {NoStop}%
\bibitem [{\citenamefont {Abelev}\ \emph {et~al.}(2013)\citenamefont {Abelev}
  \emph {et~al.}}]{PhysRevC.88.044909}%
  \BibitemOpen
  \bibfield  {author} {\bibinfo {author} {\bibfnamefont {B.}~\bibnamefont
  {Abelev}} \emph {et~al.} (\bibinfo {collaboration} {ALICE Collaboration}),\
  }\href {\doibase 10.1103/PhysRevC.88.044909} {\bibfield  {journal} {\bibinfo
  {journal} {Phys. Rev. C}\ }\textbf {\bibinfo {volume} {88}},\ \bibinfo
  {pages} {044909} (\bibinfo {year} {2013})}\BibitemShut {NoStop}%
\bibitem [{\citenamefont {Zapp}(2014)}]{Zapp2014}%
  \BibitemOpen
  \bibfield  {author} {\bibinfo {author} {\bibfnamefont {K.}~\bibnamefont
  {Zapp}},\ }\href {\doibase 10.1140/epjc/s10052-014-2762-1} {\bibfield
  {journal} {\bibinfo  {journal} {Eur.\ Phys.\ J.\ C}\ }\textbf {\bibinfo
  {volume} {74}},\ \bibinfo {pages} {2762} (\bibinfo {year}
  {2014})}\BibitemShut {NoStop}%
\bibitem [{\citenamefont {He}\ \emph {et~al.}(2015)\citenamefont {He},
  \citenamefont {Luo}, \citenamefont {Wang},\ and\ \citenamefont
  {Zhu}}]{PhysRevC.91.054908}%
  \BibitemOpen
  \bibfield  {author} {\bibinfo {author} {\bibfnamefont {Y.}~\bibnamefont
  {He}}, \bibinfo {author} {\bibfnamefont {T.}~\bibnamefont {Luo}}, \bibinfo
  {author} {\bibfnamefont {X.-N.}\ \bibnamefont {Wang}}, \ and\ \bibinfo
  {author} {\bibfnamefont {Y.}~\bibnamefont {Zhu}},\ }\href {\doibase
  10.1103/PhysRevC.91.054908} {\bibfield  {journal} {\bibinfo  {journal} {Phys.
  Rev. C}\ }\textbf {\bibinfo {volume} {91}},\ \bibinfo {pages} {054908}
  (\bibinfo {year} {2015})}\BibitemShut {NoStop}%
\bibitem [{\citenamefont {Sj{\"{o}}strand}\ \emph {et~al.}(2015)\citenamefont
  {Sj{\"{o}}strand} \emph {et~al.}}]{SJOSTRAND2015159}%
  \BibitemOpen
  \bibfield  {author} {\bibinfo {author} {\bibfnamefont {T.}~\bibnamefont
  {Sj{\"{o}}strand}} \emph {et~al.},\ }\href {\doibase
  10.1016/j.cpc.2015.01.024} {\bibfield  {journal} {\bibinfo  {journal}
  {Comput. Phys. Commun.}\ }\textbf {\bibinfo {volume} {191}},\ \bibinfo
  {pages} {159} (\bibinfo {year} {2015})}\BibitemShut {NoStop}%
\bibitem [{\citenamefont {Khachatryan}\ \emph {et~al.}(2016)\citenamefont
  {Khachatryan} \emph {et~al.}}]{Khachatryan2016}%
  \BibitemOpen
  \bibfield  {author} {\bibinfo {author} {\bibfnamefont {V.}~\bibnamefont
  {Khachatryan}} \emph {et~al.},\ }\href {\doibase
  10.1140/epjc/s10052-016-3988-x} {\bibfield  {journal} {\bibinfo  {journal}
  {Eur. Phys. J. C}\ }\textbf {\bibinfo {volume} {76}},\ \bibinfo {pages} {155}
  (\bibinfo {year} {2016})}\BibitemShut {NoStop}%
\bibitem [{\citenamefont {Gubser}\ \emph {et~al.}(2008)\citenamefont {Gubser},
  \citenamefont {Pufu},\ and\ \citenamefont {Yarom}}]{PhysRevD.78.066014}%
  \BibitemOpen
  \bibfield  {author} {\bibinfo {author} {\bibfnamefont {S.~S.}\ \bibnamefont
  {Gubser}}, \bibinfo {author} {\bibfnamefont {S.~S.}\ \bibnamefont {Pufu}}, \
  and\ \bibinfo {author} {\bibfnamefont {A.}~\bibnamefont {Yarom}},\ }\href
  {\doibase 10.1103/PhysRevD.78.066014} {\bibfield  {journal} {\bibinfo
  {journal} {Phys. Rev. D}\ }\textbf {\bibinfo {volume} {78}},\ \bibinfo
  {pages} {066014} (\bibinfo {year} {2008})}\BibitemShut {NoStop}%
\bibitem [{\citenamefont {Gubser}(2010)}]{PhysRevD.82.085027}%
  \BibitemOpen
  \bibfield  {author} {\bibinfo {author} {\bibfnamefont {S.~S.}\ \bibnamefont
  {Gubser}},\ }\href {\doibase 10.1103/PhysRevD.82.085027} {\bibfield
  {journal} {\bibinfo  {journal} {Phys. Rev. D}\ }\textbf {\bibinfo {volume}
  {82}},\ \bibinfo {pages} {085027} (\bibinfo {year} {2010})}\BibitemShut
  {NoStop}%
\bibitem [{\citenamefont {Lokhtin}\ and\ \citenamefont
  {Snigirev}(2006)}]{Lokhtin2006}%
  \BibitemOpen
  \bibfield  {author} {\bibinfo {author} {\bibfnamefont {I.~P.}\ \bibnamefont
  {Lokhtin}}\ and\ \bibinfo {author} {\bibfnamefont {A.~M.}\ \bibnamefont
  {Snigirev}},\ }\href {\doibase 10.1140/epjc/s2005-02426-3} {\bibfield
  {journal} {\bibinfo  {journal} {Eur.\ Phys.\ J.\ C}\ }\textbf {\bibinfo
  {volume} {45}},\ \bibinfo {pages} {211} (\bibinfo {year} {2006})}\BibitemShut
  {NoStop}%
\bibitem [{\citenamefont {Adam}\ \emph {et~al.}(2016)\citenamefont {Adam} \emph
  {et~al.}}]{PhysRevLett.116.222302}%
  \BibitemOpen
  \bibfield  {author} {\bibinfo {author} {\bibfnamefont {J.}~\bibnamefont
  {Adam}} \emph {et~al.} (\bibinfo {collaboration} {ALICE Collaboration}),\
  }\href {\doibase 10.1103/PhysRevLett.116.222302} {\bibfield  {journal}
  {\bibinfo  {journal} {Phys. Rev. Lett.}\ }\textbf {\bibinfo {volume} {116}},\
  \bibinfo {pages} {222302} (\bibinfo {year} {2016})}\BibitemShut {NoStop}%
\bibitem [{\citenamefont {Cacciari}\ \emph {et~al.}(2008)\citenamefont
  {Cacciari}, \citenamefont {Salam},\ and\ \citenamefont
  {Soyez}}]{1126-6708-2008-04-063}%
  \BibitemOpen
  \bibfield  {author} {\bibinfo {author} {\bibfnamefont {M.}~\bibnamefont
  {Cacciari}}, \bibinfo {author} {\bibfnamefont {G.~P.}\ \bibnamefont {Salam}},
  \ and\ \bibinfo {author} {\bibfnamefont {G.}~\bibnamefont {Soyez}},\ }\href
  {http://stacks.iop.org/1126-6708/2008/i=04/a=063} {\bibfield  {journal}
  {\bibinfo  {journal} {J. High Energy Phys.}\ }\textbf {\bibinfo {volume}
  {2008}},\ \bibinfo {pages} {063} (\bibinfo {year} {2008})}\BibitemShut
  {NoStop}%
\bibitem [{\citenamefont {Komiske}\ \emph {et~al.}(2018)\citenamefont
  {Komiske}, \citenamefont {Metodiev},\ and\ \citenamefont
  {Thaler}}]{Komiske2018}%
  \BibitemOpen
  \bibfield  {author} {\bibinfo {author} {\bibfnamefont {P.~T.}\ \bibnamefont
  {Komiske}}, \bibinfo {author} {\bibfnamefont {E.~M.}\ \bibnamefont
  {Metodiev}}, \ and\ \bibinfo {author} {\bibfnamefont {J.}~\bibnamefont
  {Thaler}},\ }\href {\doibase 10.1007/JHEP04(2018)013} {\bibfield  {journal}
  {\bibinfo  {journal} {Journal of High Energy Physics}\ }\textbf {\bibinfo
  {volume} {2018}},\ \bibinfo {pages} {13} (\bibinfo {year}
  {2018})}\BibitemShut {NoStop}%
\bibitem [{\citenamefont {Dongarra}(2002)}]{doi:10.1177/10943420020160010101}%
  \BibitemOpen
  \bibfield  {author} {\bibinfo {author} {\bibfnamefont {J.}~\bibnamefont
  {Dongarra}},\ }\href {\doibase 10.1177/10943420020160010101} {\bibfield
  {journal} {\bibinfo  {journal} {Int. J. High Perform. Comput. Appl.}\
  }\textbf {\bibinfo {volume} {16}},\ \bibinfo {pages} {1} (\bibinfo {year}
  {2002})}\BibitemShut {NoStop}%
\bibitem [{\citenamefont {Blackford}\ \emph {et~al.}(2002)\citenamefont
  {Blackford} \emph {et~al.}}]{2002:USB:567806.567807}%
  \BibitemOpen
  \bibfield  {author} {\bibinfo {author} {\bibfnamefont {L.~S.}\ \bibnamefont
  {Blackford}} \emph {et~al.},\ }\href {\doibase 10.1145/567806.567807}
  {\bibfield  {journal} {\bibinfo  {journal} {ACM Trans. Math. Softw.}\
  }\textbf {\bibinfo {volume} {28}},\ \bibinfo {pages} {135} (\bibinfo {year}
  {2002})}\BibitemShut {NoStop}%
\bibitem [{\citenamefont {Choromanska}\ \emph {et~al.}(2015)\citenamefont
  {Choromanska}, \citenamefont {Henaff}, \citenamefont {Mathieu}, \citenamefont
  {Ben~Arous},\ and\ \citenamefont {LeCun}}]{pmlr-v38-choromanska15}%
  \BibitemOpen
  \bibfield  {author} {\bibinfo {author} {\bibfnamefont {A.}~\bibnamefont
  {Choromanska}}, \bibinfo {author} {\bibfnamefont {M.}~\bibnamefont {Henaff}},
  \bibinfo {author} {\bibfnamefont {M.}~\bibnamefont {Mathieu}}, \bibinfo
  {author} {\bibfnamefont {G.}~\bibnamefont {Ben~Arous}}, \ and\ \bibinfo
  {author} {\bibfnamefont {Y.}~\bibnamefont {LeCun}},\ }in\ \href
  {http://proceedings.mlr.press/v38/choromanska15.html} {\emph {\bibinfo
  {booktitle} {Proceedings of the Eighteenth International Conference on
  Artificial Intelligence and Statistics}}},\ \bibinfo {series} {Proceedings of
  Machine Learning Research}, Vol.~\bibinfo {volume} {38},\ \bibinfo {editor}
  {edited by\ \bibinfo {editor} {\bibfnamefont {G.}~\bibnamefont {Lebanon}}\
  and\ \bibinfo {editor} {\bibfnamefont {S.~V.~N.}\ \bibnamefont
  {Vishwanathan}}}\ (\bibinfo {year} {2015})\ pp.\ \bibinfo {pages}
  {192--204}\BibitemShut {NoStop}%
\bibitem [{\citenamefont {Haeffele}\ and\ \citenamefont
  {Vidal}(2015)}]{Haeffele2015}%
  \BibitemOpen
  \bibfield  {author} {\bibinfo {author} {\bibfnamefont {B.~D.}\ \bibnamefont
  {Haeffele}}\ and\ \bibinfo {author} {\bibfnamefont {R.}~\bibnamefont
  {Vidal}},\ }\href@noop {} {\enquote {\bibinfo {title} {Global optimality in
  tensor factorization, deep learning, and beyond},}\ } (\bibinfo {year}
  {2015}),\ \Eprint {http://arxiv.org/abs/1506.07540} {arXiv:1506.07540
  [cs.NA]} \BibitemShut {NoStop}%
\bibitem [{\citenamefont {Janzamin}\ \emph {et~al.}(2015)\citenamefont
  {Janzamin}, \citenamefont {Sedghi},\ and\ \citenamefont
  {Anandkumar}}]{Janzamin2015}%
  \BibitemOpen
  \bibfield  {author} {\bibinfo {author} {\bibfnamefont {M.}~\bibnamefont
  {Janzamin}}, \bibinfo {author} {\bibfnamefont {H.}~\bibnamefont {Sedghi}}, \
  and\ \bibinfo {author} {\bibfnamefont {A.}~\bibnamefont {Anandkumar}},\
  }\href@noop {} {\enquote {\bibinfo {title} {Generalization bounds for neural
  networks through tensor factorization},}\ } (\bibinfo {year} {2015}),\
  \Eprint {http://arxiv.org/abs/1506.08473} {arXiv:1506.08473 [cs.LG]}
  \BibitemShut {NoStop}%
\bibitem [{\citenamefont {Ioffe}\ and\ \citenamefont
  {Szegedy}(2015)}]{pmlr-v37-ioffe15}%
  \BibitemOpen
  \bibfield  {author} {\bibinfo {author} {\bibfnamefont {S.}~\bibnamefont
  {Ioffe}}\ and\ \bibinfo {author} {\bibfnamefont {C.}~\bibnamefont
  {Szegedy}},\ }in\ \href {http://proceedings.mlr.press/v37/ioffe15.html}
  {\emph {\bibinfo {booktitle} {Proceedings of the 32nd International
  Conference on Machine Learning}}},\ \bibinfo {series} {Proceedings of Machine
  Learning Research}, Vol.~\bibinfo {volume} {37},\ \bibinfo {editor} {edited
  by\ \bibinfo {editor} {\bibfnamefont {F.}~\bibnamefont {Bach}}\ and\ \bibinfo
  {editor} {\bibfnamefont {D.}~\bibnamefont {Blei}}}\ (\bibinfo {year} {2015})\
  pp.\ \bibinfo {pages} {448--456}\BibitemShut {NoStop}%
\bibitem [{\citenamefont {Jiang}\ and\ \citenamefont
  {Guo}(2013)}]{doi:10.1063/1.4817187}%
  \BibitemOpen
  \bibfield  {author} {\bibinfo {author} {\bibfnamefont {B.}~\bibnamefont
  {Jiang}}\ and\ \bibinfo {author} {\bibfnamefont {H.}~\bibnamefont {Guo}},\
  }\href {\doibase 10.1063/1.4817187} {\bibfield  {journal} {\bibinfo
  {journal} {J. Chem. Phys.}\ }\textbf {\bibinfo {volume} {139}},\ \bibinfo
  {pages} {054112} (\bibinfo {year} {2013})}\BibitemShut {NoStop}%
\bibitem [{\citenamefont {Hahnloser}\ \emph {et~al.}(2000)\citenamefont
  {Hahnloser}, \citenamefont {Sarpeshkar}, \citenamefont {Mahowald},
  \citenamefont {Douglas},\ and\ \citenamefont {Seung}}]{Hahnloser2000}%
  \BibitemOpen
  \bibfield  {author} {\bibinfo {author} {\bibfnamefont {R.~H.~R.}\
  \bibnamefont {Hahnloser}}, \bibinfo {author} {\bibfnamefont {R.}~\bibnamefont
  {Sarpeshkar}}, \bibinfo {author} {\bibfnamefont {M.~A.}\ \bibnamefont
  {Mahowald}}, \bibinfo {author} {\bibfnamefont {R.~J.}\ \bibnamefont
  {Douglas}}, \ and\ \bibinfo {author} {\bibfnamefont {H.~S.}\ \bibnamefont
  {Seung}},\ }\href {\doibase 10.1038/35016072} {\bibfield  {journal} {\bibinfo
   {journal} {Nature}\ }\textbf {\bibinfo {volume} {405}},\ \bibinfo {pages}
  {947} (\bibinfo {year} {2000})}\BibitemShut {NoStop}%
\bibitem [{\citenamefont {Glorot}\ \emph {et~al.}(2011)\citenamefont {Glorot},
  \citenamefont {Bordes},\ and\ \citenamefont {Bengio}}]{pmlr-v15-glorot11a}%
  \BibitemOpen
  \bibfield  {author} {\bibinfo {author} {\bibfnamefont {X.}~\bibnamefont
  {Glorot}}, \bibinfo {author} {\bibfnamefont {A.}~\bibnamefont {Bordes}}, \
  and\ \bibinfo {author} {\bibfnamefont {Y.}~\bibnamefont {Bengio}},\ }in\
  \href {http://proceedings.mlr.press/v15/glorot11a.html} {\emph {\bibinfo
  {booktitle} {Proceedings of the Fourteenth International Conference on
  Artificial Intelligence and Statistics}}},\ \bibinfo {series} {Proceedings of
  Machine Learning Research}, Vol.~\bibinfo {volume} {15},\ \bibinfo {editor}
  {edited by\ \bibinfo {editor} {\bibfnamefont {G.}~\bibnamefont {Gordon}},
  \bibinfo {editor} {\bibfnamefont {D.}~\bibnamefont {Dunson}}, \ and\ \bibinfo
  {editor} {\bibfnamefont {M.}~\bibnamefont {Dud{\'{\i}}k}}}\ (\bibinfo {year}
  {2011})\ pp.\ \bibinfo {pages} {315--323}\BibitemShut {NoStop}%
\bibitem [{\citenamefont {Bridle}(1990)}]{10.1007/978-3-642-76153-9_28}%
  \BibitemOpen
  \bibfield  {author} {\bibinfo {author} {\bibfnamefont {J.~S.}\ \bibnamefont
  {Bridle}},\ }in\ \href {\doibase 10.1007/978-3-642-76153-9_28} {\emph
  {\bibinfo {booktitle} {Neurocomputing}}},\ \bibinfo {series} {{NATO} {ASI}
  Series {F}: Computer and Systems Sciences}, Vol.~\bibinfo {volume} {68},\
  \bibinfo {editor} {edited by\ \bibinfo {editor} {\bibfnamefont {F.~F.}\
  \bibnamefont {Souli{\'e}}}\ and\ \bibinfo {editor} {\bibfnamefont
  {J.}~\bibnamefont {H{\'e}rault}}}\ (\bibinfo  {publisher} {Springer},\
  \bibinfo {address} {Berlin, Heidelberg},\ \bibinfo {year} {1990})\ pp.\
  \bibinfo {pages} {227--236}\BibitemShut {NoStop}%
\bibitem [{\citenamefont {Krogh}\ and\ \citenamefont
  {Hertz}(1992)}]{NIPS1991_563}%
  \BibitemOpen
  \bibfield  {author} {\bibinfo {author} {\bibfnamefont {A.}~\bibnamefont
  {Krogh}}\ and\ \bibinfo {author} {\bibfnamefont {J.~A.}\ \bibnamefont
  {Hertz}},\ }in\ \href
  {http://papers.nips.cc/paper/563-a-simple-weight-decay-can-improve-generalization.pdf}
  {\emph {\bibinfo {booktitle} {Advances in Neural Information Processing
  Systems 4}}},\ \bibinfo {editor} {edited by\ \bibinfo {editor} {\bibfnamefont
  {J.~E.}\ \bibnamefont {Moody}}, \bibinfo {editor} {\bibfnamefont {S.~J.}\
  \bibnamefont {Hanson}}, \ and\ \bibinfo {editor} {\bibfnamefont {R.~P.}\
  \bibnamefont {Lippmann}}}\ (\bibinfo  {publisher} {Morgan-Kaufmann},\
  \bibinfo {year} {1992})\ pp.\ \bibinfo {pages} {950--957}\BibitemShut
  {NoStop}%
\bibitem [{\citenamefont {Cisse}\ \emph {et~al.}(2017)\citenamefont {Cisse},
  \citenamefont {Bojanowski}, \citenamefont {Grave}, \citenamefont {Dauphin},\
  and\ \citenamefont {Usunier}}]{Cisse2017}%
  \BibitemOpen
  \bibfield  {author} {\bibinfo {author} {\bibfnamefont {M.}~\bibnamefont
  {Cisse}}, \bibinfo {author} {\bibfnamefont {P.}~\bibnamefont {Bojanowski}},
  \bibinfo {author} {\bibfnamefont {E.}~\bibnamefont {Grave}}, \bibinfo
  {author} {\bibfnamefont {Y.}~\bibnamefont {Dauphin}}, \ and\ \bibinfo
  {author} {\bibfnamefont {N.}~\bibnamefont {Usunier}},\ }\href@noop {}
  {\enquote {\bibinfo {title} {{Parseval} networks: Improving robustness to
  adversarial examples},}\ } (\bibinfo {year} {2017}),\ \Eprint
  {http://arxiv.org/abs/1704.08847} {arXiv:1704.08847 [stat.ML]} \BibitemShut
  {NoStop}%
\bibitem [{\citenamefont {Gouk}\ \emph {et~al.}(2018)\citenamefont {Gouk},
  \citenamefont {Frank}, \citenamefont {Pfahringer},\ and\ \citenamefont
  {Cree}}]{Gouk2018}%
  \BibitemOpen
  \bibfield  {author} {\bibinfo {author} {\bibfnamefont {H.}~\bibnamefont
  {Gouk}}, \bibinfo {author} {\bibfnamefont {E.}~\bibnamefont {Frank}},
  \bibinfo {author} {\bibfnamefont {B.}~\bibnamefont {Pfahringer}}, \ and\
  \bibinfo {author} {\bibfnamefont {M.}~\bibnamefont {Cree}},\ }\href@noop {}
  {\enquote {\bibinfo {title} {Regularisation of neural networks by enforcing
  {Lipschitz} continuity},}\ } (\bibinfo {year} {2018}),\ \Eprint
  {http://arxiv.org/abs/1804.04368} {arXiv:1804.04368 [stat.ML]} \BibitemShut
  {NoStop}%
\bibitem [{\citenamefont {Santosa}\ and\ \citenamefont
  {Symes}(1986)}]{doi:10.1137/0907087}%
  \BibitemOpen
  \bibfield  {author} {\bibinfo {author} {\bibfnamefont {F.}~\bibnamefont
  {Santosa}}\ and\ \bibinfo {author} {\bibfnamefont {W.~W.}\ \bibnamefont
  {Symes}},\ }\href {\doibase 10.1137/0907087} {\bibfield  {journal} {\bibinfo
  {journal} {SIAM J. Sci. Stat. Comput.}\ }\textbf {\bibinfo {volume} {7}},\
  \bibinfo {pages} {1307} (\bibinfo {year} {1986})}\BibitemShut {NoStop}%
\bibitem [{\citenamefont {Tibshirani}(1996)}]{10.2307/2346178}%
  \BibitemOpen
  \bibfield  {author} {\bibinfo {author} {\bibfnamefont {R.}~\bibnamefont
  {Tibshirani}},\ }\href {http://www.jstor.org/stable/2346178} {\bibfield
  {journal} {\bibinfo  {journal} {J. Royal Stat. Soc. B}\ }\textbf {\bibinfo
  {volume} {58}},\ \bibinfo {pages} {267} (\bibinfo {year} {1996})}\BibitemShut
  {NoStop}%
\bibitem [{\citenamefont {Abadi}\ \emph {et~al.}(2016)\citenamefont {Abadi}
  \emph {et~al.}}]{tensorflow2015-whitepaper}%
  \BibitemOpen
  \bibfield  {author} {\bibinfo {author} {\bibfnamefont {M.}~\bibnamefont
  {Abadi}} \emph {et~al.},\ }in\ \href
  {https://www.usenix.org/conference/osdi16/technical-sessions/presentation/abadi}
  {\emph {\bibinfo {booktitle} {Proceedings of OSDI'16: 12th USENIX Symposium
  on Operating Systems Design and Implementation}}}\ (\bibinfo {organization}
  {USENIX Association},\ \bibinfo {address} {Berkeley, CA},\ \bibinfo {year}
  {2016})\ pp.\ \bibinfo {pages} {265--283}\BibitemShut {NoStop}%
\bibitem [{\citenamefont {Chetlur}\ \emph {et~al.}(2014)\citenamefont
  {Chetlur}, \citenamefont {Woolley}, \citenamefont {Vandermersch},
  \citenamefont {Cohen}, \citenamefont {Tran}, \citenamefont {Catanzaro},\ and\
  \citenamefont {Shelhamer}}]{ChetlurWVCTCS14}%
  \BibitemOpen
  \bibfield  {author} {\bibinfo {author} {\bibfnamefont {S.}~\bibnamefont
  {Chetlur}}, \bibinfo {author} {\bibfnamefont {C.}~\bibnamefont {Woolley}},
  \bibinfo {author} {\bibfnamefont {P.}~\bibnamefont {Vandermersch}}, \bibinfo
  {author} {\bibfnamefont {J.}~\bibnamefont {Cohen}}, \bibinfo {author}
  {\bibfnamefont {J.}~\bibnamefont {Tran}}, \bibinfo {author} {\bibfnamefont
  {B.}~\bibnamefont {Catanzaro}}, \ and\ \bibinfo {author} {\bibfnamefont
  {E.}~\bibnamefont {Shelhamer}},\ }\href@noop {} {\enquote {\bibinfo {title}
  {{cuDNN:} efficient primitives for deep learning},}\ } (\bibinfo {year}
  {2014}),\ \Eprint {http://arxiv.org/abs/1410.0759} {arXiv:1410.0759 [cs.NE]}
  \BibitemShut {NoStop}%
\bibitem [{\citenamefont {McConaghy}(2011{\natexlab{a}})}]{McConaghy2011}%
  \BibitemOpen
  \bibfield  {author} {\bibinfo {author} {\bibfnamefont {T.}~\bibnamefont
  {McConaghy}},\ }\enquote {\bibinfo {title} {{FFX:} fast, scalable,
  deterministic symbolic regression technology},}\ in\ \href {\doibase
  10.1007/978-1-4614-1770-5_13} {\emph {\bibinfo {booktitle} {Genetic
  Programming Theory and Practice IX}}},\ \bibinfo {editor} {edited by\
  \bibinfo {editor} {\bibfnamefont {R.}~\bibnamefont {Riolo}}, \bibinfo
  {editor} {\bibfnamefont {E.}~\bibnamefont {Vladislavleva}}, \ and\ \bibinfo
  {editor} {\bibfnamefont {J.~H.}\ \bibnamefont {Moore}}}\ (\bibinfo
  {publisher} {Springer},\ \bibinfo {address} {New York, NY},\ \bibinfo {year}
  {2011})\ pp.\ \bibinfo {pages} {235--260}\BibitemShut {NoStop}%
\bibitem [{\citenamefont {McConaghy}(2011{\natexlab{b}})}]{6055329}%
  \BibitemOpen
  \bibfield  {author} {\bibinfo {author} {\bibfnamefont {T.}~\bibnamefont
  {McConaghy}},\ }in\ \href {\doibase 10.1109/CICC.2011.6055329} {\emph
  {\bibinfo {booktitle} {2011 IEEE Custom Integrated Circuits Conference
  (CICC)}}}\ (\bibinfo  {publisher} {IEEE},\ \bibinfo {address} {Piscataway,
  NJ},\ \bibinfo {year} {2011})\ pp.\ \bibinfo {pages} {1--8}\BibitemShut
  {NoStop}%
\bibitem [{\citenamefont {Kingma}\ and\ \citenamefont {Ba}(2014)}]{KingmaB14}%
  \BibitemOpen
  \bibfield  {author} {\bibinfo {author} {\bibfnamefont {D.~P.}\ \bibnamefont
  {Kingma}}\ and\ \bibinfo {author} {\bibfnamefont {J.}~\bibnamefont {Ba}},\
  }\href@noop {} {\enquote {\bibinfo {title} {Adam: {A} method for stochastic
  optimization},}\ } (\bibinfo {year} {2014}),\ \Eprint
  {http://arxiv.org/abs/1412.6980} {arXiv:1412.6980 [cs.LG]} \BibitemShut
  {NoStop}%
\bibitem [{\citenamefont {B{\"a}hr}\ \emph {et~al.}(2008)\citenamefont
  {B{\"a}hr} \emph {et~al.}}]{Baehr2008}%
  \BibitemOpen
  \bibfield  {author} {\bibinfo {author} {\bibfnamefont {M.}~\bibnamefont
  {B{\"a}hr}} \emph {et~al.},\ }\href {\doibase 10.1140/epjc/s10052-008-0798-9}
  {\bibfield  {journal} {\bibinfo  {journal} {The European Physical Journal C}\
  }\textbf {\bibinfo {volume} {58}},\ \bibinfo {pages} {639} (\bibinfo {year}
  {2008})}\BibitemShut {NoStop}%
\bibitem [{\citenamefont {Bellm}\ \emph {et~al.}(2016)\citenamefont {Bellm}
  \emph {et~al.}}]{Bellm2016}%
  \BibitemOpen
  \bibfield  {author} {\bibinfo {author} {\bibfnamefont {J.}~\bibnamefont
  {Bellm}} \emph {et~al.},\ }\href {\doibase 10.1140/epjc/s10052-016-4018-8}
  {\bibfield  {journal} {\bibinfo  {journal} {The European Physical Journal C}\
  }\textbf {\bibinfo {volume} {76}},\ \bibinfo {pages} {196} (\bibinfo {year}
  {2016})}\BibitemShut {NoStop}%
\bibitem [{\citenamefont {Sirunyan}\ \emph
  {et~al.}(2018{\natexlab{b}})\citenamefont {Sirunyan} \emph
  {et~al.}}]{Sirunyan:2018gct}%
  \BibitemOpen
  \bibfield  {author} {\bibinfo {author} {\bibfnamefont {A.~M.}\ \bibnamefont
  {Sirunyan}} \emph {et~al.} (\bibinfo {collaboration} {CMS}),\ }\href@noop {}
  {\  (\bibinfo {year} {2018}{\natexlab{b}})},\ \Eprint
  {http://arxiv.org/abs/1805.05145} {arXiv:1805.05145 [hep-ex]} \BibitemShut
  {NoStop}%
\end{thebibliography}%

\end{document}